\documentclass{article}
\usepackage[utf8]{inputenc}
\usepackage{amsmath,amsthm, amssymb, amsfonts}
\usepackage{fullpage}

\usepackage{wasysym}
\usepackage{todonotes}
\usepackage{comment}
\usepackage{mathtools}
\usepackage{wrapfig}
\usepackage{microtype}
\usepackage[algo2e]{algorithm2e}
\usepackage{xspace}
\usepackage{svg}

\usepackage{algorithm,algorithmic}

\definecolor{darkgreen}{rgb}{0,0.5,0}
\definecolor{darkblue}{RGB}{37,70,160}
\definecolor{amber}{rgb}{1.0, 0.75, 0.0}
\definecolor{aquamarine}{rgb}{0.4, 0.8, 0.66}
\definecolor{purple}{RGB}{186, 121, 246}
\definecolor{green}{rgb}{0.2,0.6,0.15}

\usepackage{tikz,xcolor}
\usepackage{tikz-3dplot}
\usetikzlibrary{3d,decorations.text,shapes.arrows,positioning,fit,backgrounds}
\usepackage{pgflibraryplotmarks}
\usepackage{pgflibrarysnakes}
\usetikzlibrary{calc,fadings,snakes,matrix,positioning,spy,decorations.pathreplacing,decorations.text}

\tikzset{pics/fake box/.style args={
#1 with dimensions #2 and #3 and #4}{code={\draw[gray,ultra thin,fill=#1]  (0,0,0) coordinate(-front-bottom-left) to
++ (0,#3,0) coordinate(-front-top-right) --++
(#2,0,0) coordinate(-front-top-right) --++ (0,-#3,0)
coordinate(-front-bottom-right) -- cycle;
\draw[gray,ultra thin,fill=#1] (0,#3,0)  --++
 (0,0,#4) coordinate(-back-top-left) --++ (#2,0,0)
 coordinate(-back-top-right) --++ (0,0,-#4)  -- cycle;
\draw[gray,ultra thin,fill=#1!80!black] (#2,0,0) --++ (0,0,#4) coordinate(-back-bottom-right)
--++ (0,#3,0) --++ (0,0,-#4) -- cycle;
\path[gray,decorate,decoration={text effects along path,text={CONV}}] (#2/2,{2+(#3-2)/2},0) -- (#2/2,0,0);
}
}}

\tikzset{circle dotted/.style={dash pattern=on 0.05mm off 2mm,line cap=round}}

\newcommand{\PW}{\mathcal{PW}}

\newcommand{\Natural}{\mathbb{N}}

\newcommand{\Real}{\mathbb{R}}

\newcommand{\supp}{\operatorname{supp}}
\newcommand{\ssupp}{\operatorname{sing\,supp}}
\newcommand{\DWF}{\operatorname{DWF}}
\newcommand{\WF}{\operatorname{WF}}
\newcommand{\FBP}{\operatorname{FBP}}
\newcommand{\differential}{\mathrm{d}}

\newcommand{\datamanifold}{\mathbb{M}}

\newtheorem{theorem}{Theorem}[section]
\newtheorem*{theorem*}{Theorem}

\newtheorem{definition}[theorem]{Definition}

\newtheorem*{remark*}{Remark}
\newtheorem*{proposition*}{Proposition}

\numberwithin{equation}{section}

\definecolor{darkcandyapplered}{rgb}{0.64, 0.0, 0.0}

\title{Shearlets as Feature Extractor for Semantic Edge Detection: The Model-Based and Data-Driven Realm}

\author{H\'ector Andrade-Loarca$^{1}$, Gitta Kutyniok$^{1,2,3}$ and Ozan \"Oktem$^{4}$}
 
\date{}

\begin{document}
\maketitle

\footnotetext[1]{Institut f\"ur Mathematik, Technische Universit\"at Berlin, 10623 Berlin, Germany, \texttt{$\{$kutyniok,andrade$\}$@math.tu-berlin.de}}
\footnotetext[2]{Fakult\"at Elektrotechnik und Informatik, 10587 Berlin, Germany, \texttt{kutyniok@math.tu-berlin.de}}
\footnotetext[3]{Department of Physics and Technology, University of Troms\o}
\footnotetext[4]{Department of Mathematics, KTH - Royal Institute of Technology,
SE-100 44 Stockholm, Sweden, \texttt{ozan@kth.se}}

\begin{abstract}
Semantic edge detection has recently gained a lot of attention as an image processing task, mainly due to its wide range of real-world applications. This is based on the fact that edges in images contain most of the semantic information. Semantic edge detection involves two tasks, namely pure edge detecion and edge classification. Those are in fact fundamentally distinct in terms of the level of abstraction that each task requires, which is known as the distracted supervision paradox that limits the possible performance of a supervised model in semantic edge detection.  In this work, we will present a novel hybrid method to avoid the distracted supervision paradox and achieve high-performance in semantic edge detection. Our approach is based on a combination of the model-based concept of shearlets, which provides probably optimally sparse approximations of a model-class of images, and the data-driven method of a suitably designed convolutional neural netwok. Finally, we present several applications such as tomographic reconstruction and show that our approach signifiantly outperforms former methods, thereby indicating 
the value of such hybrid methods for the area in biomedical imaging.
\end{abstract}

\medskip
\noindent
\textbf{Keywords:} Multiscale geometric analysis, Harmonic analysis, Deep learning, Feature extraction.
\smallskip

\noindent
\textbf{Mathematics Subject Classification:} 42Bxx, 35A18, 65T60, 68T10.

\section{Introduction}
\label{sec:Intro}

In computer vision, semantic edge detection is the task of detecting edges and object boundaries in natural images and classifying the points in those edges from a finite set of classes, which, for instance, represent the objects the edges belong to \cite{yu2017casenet} or the orientation of the edge at that particular point \cite{andrade2019wfset}. The recent interest from the research community in semantic edge detection is mainly driven by its far-reaching applications in imaging related tasks such as object recognition, semantic segmentation, and image reconstruction. 

Semantic edge detection combines two different classification tasks. The first is classical \emph{category-agnostic} edge detection, which can be viewed as a pixel-wise binary classification for determining whether a pixel belongs to an edge or not. The second is the recognition of the classes of pixels in an image that belong to edges.

One can perform semantic edge detection using a model-based or a data-driven approach; each comes with their strengths and shortcomings. The main idea presented here is to overcome the shortcomings by combining elements of model-based and data-driven approaches for semantic edge detection. We remark that this objective follows a common thread of current research, namely aiming for an optimal combination of model-based or a data-driven methodologies.

Our conceptually general approach aims to combine advances from several fields. First, it utilizes properties that the representation systems of shearlets has, foremost to optimally represent singularities in signals \cite{gitta2005shearlets}.  Second, it also takes advantage of microlocal analysis for describing how singularities transform when acted upon by a wide range of operators \cite{Krishnan:2015aa}. Finally, it in addition leverages on the proven track record that convolutional neural networks has had in image classification \cite{lecun1998lenet}.

\subsection{Model-Based Semantic Edge Detection}
Many approaches to identify singularities in images are model-based.
These methods usually involve two steps: a filtering step to enhance edge-like features and a classification step to identify pixels belonging to edges. 

The aforementioned features are extracted using simple rules and heuristics, e.g., convolution with local difference filters correspond to operating on the image with Roberts \cite{roberts1963machine}, Sobel \cite{sobel2014edge}, and Prewitt \cite{prewitt1970object} operators. In a similar manner, the well-known Canny edge detector \cite{canny1986computational} corresponds to convolving the image with a Gaussian kernel to further identify those pixels where the gradient is high.

There have been attempts in the past to also model the semantic information of detected edges as in \cite{hariharan2011semantic}. This work was in fact among the first publications to propose a principled way to combine generic object detectors with bottom-up contours for semantic edge detection. 

Determining the orientation of an edge is particularly important in inverse problems, since this information is essential in relating edges in data to those in the signal \cite{Krishnan:2015aa}. Wavefront set extraction refers to semantic edge detection, where the classification of the edges is based on their orientation. The continuous theory of wavefront set resolution via multiscale directional systems (e.g., shearlets \cite{kutyniok2009resolution}) allows one to design model-based approaches for wavefront set extraction. These are essentially a digital implementation of the continuous theory, which filters the images before performing the corresponding classification. In Section~\ref{sec:ShearWFsets}, the concept of a wavefront set will be formally defined as well as the ability of shearlets to resolve it. 

An example of such an approach is the shearlet-based algorithm in \cite{yi2009shearlet}, which uses digital shearlets to filter an image in order to highlight the features corresponding to different orientations and scales. One then performs a simple clustering classification algorithm to classify the corresponding directions. A more recent approach is \cite{rafael2015coshrem, rafael2019molecules}, where a general directional system, known as symmetric molecules, is used to filter the directional features of images to then classify them to be edge, ridge, or blob.

These model-based approaches for semantic edge detection rely on `first principles' from approximation theory and are easy to interpret, hence can also easier to improve upon. On the other hand, the use of rigid heuristics regarding the characterization of singularities makes it difficult to utilize these methods in real world application, where the data represents empirically defined function classes. 

\subsection{Data-Driven Semantic Edge Detection}
\label{subsec:IntroDataDriveEdge}
More recently, as part of the success stories of machine learning and its successes in addressing various tasks in modern computer vision, a set of deep neural network architectures for semantic edge detection \cite{yu2017casenet, bertasius2015deepedge, yun2019sed, yu2018simultaneous} have appeared. One needs to stress that these have set a new state-of-the-art of semantic edge detection. 

In broad terms, these methods use similar principles as the model-based ones, i.e., learning filters using convolutional layers and subsequently classifying the corresponding edge pixels by sigmoid or softmax classifiers. 
Since each convolutional layer represents a level of abstraction of the features in the target images, the initial layers will represent `simple' edges, whereas deeper layers will represent more `complex' features. In that sense, the two steps involved in semantic edge detection, which are semantic-agnostic edge detection and classification followed by edge classification, are conceptually far from each other. Therefore, there is no straightforward way of jointly  learning how to extract and classify the edges. This limitation in semantic edge detection is known as \emph{the distracted supervision paradox} \cite{yun2019sed}. 

\subsubsection{The Distracted Supervision Paradox}
As mentioned before, semantic-agnostic edge detection involves locating fine detailed edges by capturing discontinuities among image regions. This makes mainly use of low level features, whereas edge classification requires identifying high-level semantics by summarizing different appearance variations of the target classes. This distracted supervision paradox prevents state-of-the-art end-to-end semantic edge detection method based on deep supervision. 

On the other hand, one could directly use high-level convolution features for semantic classification and low-level convolutional ones for non-semantic edge detection by jointly optimizing over the two corresponding losses. However, such a straightforward approach was in \cite{yu2017casenet} shown to actually decrease the performance compared with directly learning semantic edges without deep supervision.

Given the seriousness of this limitation, researchers have tried non-standard ways to avoid the distracted supervision paradox. In \cite{yu2017casenet} the authors propose the CASENet architecture based on a concatenation of convolutional residual networks. They used a new skip-layer architecture, where category-wise edge activations at the top convolutional layers are shared and fused with the same set of bottom layer features. Training such a network with a multi-label loss function on the fused activations leads to state-of-the-art on semantic edge detection. In this case, the low-level features are used to augment top classifications. 
More recently, a new training approach for CASENet was proposed in \cite{yu2018simultaneous} that is referred to as the \emph{simultaneous edge alignment and learning} (SEAL). Here, one simultaneously aligns ground truth edges and learns semantic edge detectors. Inspired by these approaches, \cite{yun2019sed} proposed an end-to-end architecture using convolutional residual layers followed by an information converter layer to transform information coming from low-level features and change it into different representations. This is needed for training category-agnostic edge detection and semantic edge classification, respectively. It allows for a single deep convolutional neural network backbone representation while outperforming CASENet and SEAL. The approach based on using information converter is known as \emph{diverse deep supervision}.

The main drawback of these approaches lies in the complexity of the related deep neural network architectures, which represents an elaborate way to avoid the distracted supervision paradox.
Furthermore, the large number of network parameters makes those methods slow and difficult to train. 

A new approach, based on a novel architecture DeNSE for wavefront set extraction was recently introduced in \cite{andrade2019wfset}. In this work the digital shearlet transform is used to map the input image to a representation that is well adapted for edge detection and further classification. The power of multiscale directional system for representing singularities allows one to use a significantly smaller architecture to obtain state-of-the-art results in semantic edge detection.
Furthermore, if the specific semantic edge detection task has $N$ target classes, the full problem can be seen as a multi-label classification class with $N+1$ target classes with the extra class corresponding to the class of edge pixels. In addition to the heavy lifting performed by the shearlet transform on the classification task by offering such a convenient representation of edges, the DeNSE architecture also separates the multi-label classification task into $N+1$ individual binary classifiers. 

This strategy allows the network to achieve high accuracy, since binary classification is significantly easier than the multi-label classification. In addition, the separation of the multi-label classification problem into small binary classifiers avoids the model to encounter the distracted supervision paradox. The edge detection and the edge classification, which originally needed different levels of abstraction, are now learned independently. DeNSE was for these reasons able to outperform all other available semantic edge detection methods where the classes were the orientations of the edges. In Section~\ref{subsec:ShearWFsetsParad} we present a way to avoid the distracted supervision paradox in the case of wavefront set extraction.  Moreover, this concept is extended to general semantic edge detection in Section~\ref{sec:Gensemantic edge detectionShearDS}.

We will however first introduce the important ideas that motivate using multiscale directional representations to sense a signal on different scales and directions. This will also show how these representations optimally represent multidimensional data such as images by extracting information on their singularities. This will provide the necessary background in understanding how multiscale directional representations, like shearlets, help in reducing the complexity of edge detection and classification.

\subsection{Multiscale Directional Systems as Feature Extractors}
Multiscale systems play an important role in applied and computational harmonic analysis for the analysis of multiscale features of a signal. In the case of 1D signals, wavelets and similar systems have been used to detect the singular support \cite{mallat1992singularities}. 
This is achieved by studying the asymptotic behavior of the corresponding localized coefficients as the scale converges to zero. Rapid decay then translates to regularity properties of the function itself. 

One can in this sense regard the wavelet transform as a feature extractor that is well suited for detecting singularities. The feature vector would then be the wavelet coefficients.
However, singularities of 2D signals are not necessarily point-like. In fact, they may very well be curve-like structures (curvilinear singularities). Wavelet type expansions are, due to their isotropic nature, therefore less suitable for representing such singularities. 

\paragraph{Geometric multiscale analysis.}
The area of geometric multiscale analysis can be regarded as a sub-field of applied and computational harmonic analysis for the study of directional multiscale systems, which address the aforementioned limitation of wavelets.
Among the first examples are curvelets \cite{candes2004curvelets}, which use parabolic (non-isotropic) scaling and rotation to efficiently represent curvilinear singularities. 

Curvelets represented a breakthrough in the optimal approximation of curvilinear singularities --- formalized as the model of cartoon-like functions --- and can be seen as the starting point of geometric multiscale analysis. However, the system comes with a major practical drawback. The rotation operator used for varying the orientation of the coefficients in the system does not have a digitization that results in an implementation, which is consistent with the continuous theory \cite{candes2006discrete}. 

One main reason for the development of the  \emph{shearlet} system \cite{gitta2005shearlets} was in fact to overcome the aforementioned limitation associated with curvelets. shearlets use the shearing operator --- instead of the rotation operator --- to vary the orientation of the coefficients. This allows for a consistent and faithful digitization. In this sense shearlets provide a uniform concept for the continuum and digital realm. 
 
The implementation of shearlets indeed inherits most of the theoretical properties of the continuous transform, including the optimal representation of the model class of cartoon-like functions \cite{kutyniok2011compact}. 
 Furthermore, there even exists a set of shearlet generators with compact support that also provide optimally sparse approximations of cartoon-like functions, thereby allowing high spatial localization \cite{Kutyniok:2016:SFD:2888419.2740960}. 
Similar to regarding the wavelet transform as a feature extractor for point singularities, the shearlet transform can be used to extract information relevant for curvilinear singularities detection, like edges in images \cite{kutyniok2017classification}.
At the same time, shearlets do allow the computation of the orientation of curvilinear singularities as well. This establishes a connection to the concept of microlocal wavefront sets. One can in fact prove that the microlocal wavefront set 
can be characterised through the asymptotic decay of the shearlet coefficients \cite{kutyniok2009resolution}. Fixing the shearing and location parameters amounts to characterizing the wavefront set at a specific scale (resolution).The formal definition of continuous shearlets and their connection to the wavefront set is given in Section~\ref{subsec:ShearWFsetsCont}.

\paragraph{Digital signals.}
Once an analog signal represented by a real valued function --- one can think of an image --- is digitized, it is not meaningful to regard its singularities in the sense of lack of regularity in analysis. For such a signal, the notion of singularity refers to an abrupt change in intensity values, which in imaging applications typically corresponds to the presence of an edge. It is indeed challenging to define a precise notion of a digital edge extractor that is practically usable and consistent with a corresponding continuous version \cite{andrade2019wfset}. Despite this problem, it is still possible to approximate a singularity detector in the digital case, by using diverse heuristics. An approach based on the shearlet transform of an image is presented in detail in Section~\ref{subsec:ShearWFsetsDig}, see also \cite{andrade2019wfset}.

In general, deep learning models employed for imaging tasks --- including those for semantic edge detection --- which involve convolutional layers can be divided in different separate stages, where feature extraction step is done at first using convolutional filters playing the role of dictionary elements, to transform the data in a particular feature vector which is convenient for the specific task. In most of the cases, the harder part of the whole process is precisely the feature extraction containing most of model parameters.

It is conceivable that applying a specific, analytically defined, feature extraction such as the shearlet transform as a pre-processing step can benefit the learning process by reducing the complexity of the overall task. Intuitively, such as pre-processing should perform most of the heavy lifting and reducing the number of parameters to be learned. This idea was applied to semantic edge extraction for the case of wavefront set extraction (see \cite{andrade2019wfset}). But it seems natural to extend this concept to other tasks of semantic edge extraction, and we will cover this idea carefully in the next sections.

\subsection{Applications of Semantic Edge Detection}
The concept of edge detection is one of the principal problems in the field of image processing and computer vision. This is due to the fact that edges in images represent boundaries of objects and carry most of the information of the associated physical scene \cite{4767769, marr1980theory, BINFORD1981205, brady1982computational}. 
It has also been shown in \cite{marr1980theory} that the human visual cortex performs multiple operations of image processing, the first of which is rough sketching involving edge detection in order to reduce the amount of information that needs to be fully acquired having that the visual cortex is sparsely connected.

Based on its important role in information processing for imaging, solely edge detection is utilized in a wide range of applications in computer vision and other fields. In particular, detecting edges allows to track objects within different frames of a video by tracking the points in their boundaries. This can, for instance, be used for object depth estimation \cite{andrade2017lf} an 3D image reconstruction. 

Moreover, semantic edge detection has a deeper impact in computer vision than solely edge extraction. If the edge classes are defined by the object the corresponding edge belongs to, one can, in particular, perform object proposal generation \cite{bertasius2015deepedge}, occlusion reasoning \cite{amer2015occ}, object detection \cite{ferrari2008objdet,ferrari2010objdet}, and image-based localization \cite{ramalingam2010local}. 

In the special case when the performed semantic edge detection includes the extraction of the wavefront set as well, the acquired semantic edges can be used to solve a wide range of inverse problems. One particular class of examples are inverse problems whose forward operator is a Fourier integral operator, which are operators arising from biomedical imaging problems such as computed tomography (CT) and magnetic resonance imaging (MRI). Section~\ref{subsec:NumResAppCT} is dedicated to an application in inverse problems of special interest in biomedical imaging, namely CT image reconstruction. To solve this, we exploit the fact that the X-ray transform is a Fourier integral operator, so the wavefront set of the image is prescribed by the wavefront set of the projected data. We refer to Section~\ref{subsec:MicroLocCT} for more details. 

\subsection{Contributions}
The main contributions of our work are three-fold:
\begin{itemize}
    \item \textbf{Hybrid semantic edge detection} by using the carefully designed model-based shearlet transform to perform semantic edge detection on a convenient image representation for directional edge extraction. This concept is first introduced for the particular case of wavefront set extraction (Section~\ref{sec:ShearWFsets}) and then extended to general semantic edge detection (Section~\ref{sec:Gensemantic edge detectionShearDS}).
    \item \textbf{Distracted supervision paradox avoidance} by splitting the $N$-classes semantic edge detection into $N+1$ individual classifiers (Section~\ref{sec:ShearWFsets}\ref{subsec:ShearWFsetsParad}).
    \item \textbf{Applications of semantic edge detection to inverse problems} by introducing the micro-canonical relation that prescribes the wavefront set of an image from the wavefront set of its transformation under a forward model. We, in particular, apply this to the inverse problem of Computed Tomography (CT)  (Section~\ref{sec:NumResApp}).
\end{itemize}

In order to evaluate our novel method, we present benchmarks on the human-annotated Semantic Boundaries Dataset (SBD) \cite{hariharan2011semcontours} and a custom dataset made of random ellipses and analytically defined semantic edges.

\section{Microlocal Analysis}
\label{sec:MicroLoc}

\subsection{Basic Definitions and Properties of Distributions}
Distributions appear in physics in various forms, e.g., the Dirac $\delta$-function is widely used for describing the density of a point mass. The $\delta$-function is described as a positive function with support at a single point that integrates to 1. It is however easy to see that there are no functions with these properties. Still, one can compute with such function and get reasonable results.
Next, distributions also arise in mathematics and in particular so in the theory of partial differential equations (PDEs). For example, fundamental solutions to PDEs are usually singular distributions and the behaviour of the singularities of these distributions encodes the behaviour of the solutions. The aim of distribution theory is to formalize the definition and mathematical calculus of distributions. In the following, we outline the basic parts of distribution theory that are necessary for microlocal analysis. 

Let $\Omega \subset \Real^n$ be an open set and $\mathcal{E}(\Omega) :=C^{\infty}(\Omega)$ is the vector space of real (or complex) valued smooth functions on $\Omega$.
Next, $\mathcal{D}(\Omega) := C^{\infty}_0(\Omega)$ is the corresponding set of smooth functions that are compactly supported in $\Omega$. 
Distributions are now continuous linear functionals (with the weak-* topology) on these function spaces. 

A more precise definition requires describing the topologies for $\mathcal{E}(\Omega)$ and $\mathcal{D}(\Omega)$.
The space $\mathcal{E}(\Omega)$ is a Fr\'echet space with the family of semi-norms
\begin{equation}\label{eq:SemiNorms} 
  d_{\alpha,K}(f) := \sup_{x \in K} \bigl\vert \partial^{\alpha} f(x) \bigr\vert
  \quad\text{where $K$ runs over all compact subsets $K \subset \Omega$.}
\end{equation}
Hence, a sequence of functions $f_n$ converges to $f$ in $\mathcal{E}(\Omega)$ if and only if all its derivatives converge uniformly on compact subsets. In particular, $\mathcal{E}(\Omega)$ is Fr\'echet space.
The topology on $\mathcal{D}(\Omega)$ is more complicated to define. A sequence $f_n$ converges in $\mathcal{D}(\Omega)$ to $f$ if there exists a compact set $K$ such that $\supp f_n  \subset  K$ ,$\supp f  \subset  K$, and all derivatives of $f_n$ converge uniformly in $K$.
\begin{definition}
The set $\mathcal{E}'(\Omega)$ is the topological dual of $\mathcal{E}(\Omega)$, i.e., the space of continuous linear functionals on $\mathcal{E}(\Omega)$. Elements in $\mathcal{E}'(\Omega)$ are called \emph{compactly supported distributions} on $\Omega$.
Likewise, the set $\mathcal{D}'(\Omega)$ is the topological dual of $\mathcal{D}'(\Omega)$, i.e., the space of continuous linear functionals on $\mathcal{D}(\Omega)$. Elements in $\mathcal{D}'(\Omega)$ are called \emph{distributions} on $\Omega$.
\end{definition}
Next, to define the support of a distribution we first observe that if $\Omega_0 \subset \Omega$ is an open subset, then $\mathcal{D}(\Omega_0)$ is a closed subspace of $\mathcal{D}(\Omega)$. Furthermore, there is a natural restriction map $\mathcal{D}'(\Omega) \to \mathcal{D}'(\Omega_0)$ for any open subset. 
This ensures that the following definition is well-defined.
\begin{definition}
Let $f \in \mathcal{D}'(\Omega)$. The support $\supp f$ of $f$ is the smallest closed set $K$ such that the restriction of $f$ to $\Omega \setminus K$ is $0$.
\end{definition}

Distributions are frequently called `generalised functions' and the next example motivates this alternative terminology.   
If $f \in L_{\text{loc}}^1(\Omega)$ (a locally integrable function), then $f$ is a distribution with the standard definition
\[ f(u) := \int_{\Omega} u(x)f(x)\,dx
    \quad\text{for $u \in \mathcal{D}(\Omega)$.}
\]    
The map $L_{\text{loc}}^1(\Omega) \to \mathcal{D}'(\Omega)$ is injective, which means that $f$ is almost everywhere determined by the distribution. 
In particular, every smooth function defines a distribution and the support of a function as a distribution coincides with its support as a function. Thus, we obtain the following inclusions:
\[ 
  \mathcal{D}(\Omega) \subset \mathcal{E}'(\Omega) \subset \mathcal{D}'(\Omega)
  \quad\text{and}\quad
  \mathcal{E}(\Omega) \subset \mathcal{D}'(\Omega).
\]
The most common example of a distribution that is not a function is the Dirac $\delta$-distribution $\delta_{x_0}$ at a point $x_0 \in\Omega$. It is defined by $\delta_{x_0} \colon \mathcal{D}(\Omega) \to \Real$  as
\[  \delta_{x_0}(f) = \langle \delta_{x_0}, f \rangle = f(x_0). \]
This is a distribution with support and singular support equal to $\{x_0\}$.

Next, we define derivatives of distributions. From the definition, it is clear that any distribution can be arbitrarily often differentiated and the result will again be a distribution. For example, any function in $L_{\text{loc}}^1(\Omega)$ has distributional derivatives of any order. 
Thus, distribution theory can be thought of as the completion of differential calculus, similar to how Lebesgue integration theory is a  completion of integral calculus. 
\begin{definition}
The partial derivative $\partial^{\alpha} u$ of $u \in \mathcal{D}'(\Omega)$ is defined by
\[ (\partial^{\alpha} u)(f) := (-1)^{\vert \alpha\vert} u\bigl( \partial^{\alpha} f \bigr)
   \quad\text{for $f \in \mathcal{D}(\Omega)$.}
\]   
\end{definition}
Finally, to extend the notion of a Fourier transform to distributions, it is convenient to introduce a third space of distributions. 
This is the space of Schwartz distributions $\mathcal{S}'(\Real^n)$ (or tempered distributions) that is defined as the topological dual of the space $\mathcal{S}(\Real^n)$, which are smooth functions in $\Real^n$ where the following semi-norms are finite:
\begin{equation}\label{eq:SemiNorms2} 
  d_{\alpha,\beta}(f) := \sup_{x} \bigl\vert x^{\alpha} \partial^{\beta} f(x) \bigr\vert
\end{equation}
This space is a Fr\'echet space and tempered distributions is the dual of $\mathcal{S}(\Real^n)$.
One can now define the Fourier transform of a tempered distribution $u \in \mathcal{S}'(\Real^n)$ as
\[ 
  \widehat{u}(f) := u(\widehat{f}) \quad\text{for $f \in \mathcal{S}(\Real^n)$.}
\]
By duality (using the Plancherel formula) one can show that the Fourier transform with the above definition extends to a weak-* continuous linear map from $\mathcal{S}'(\Real^n)$ to $\mathcal{S}'(\Real^n)$.

As a final note, the above constructions extend to functions on manifolds.  
Microlocal analysis deals with the detailed analysis of how singularities of distributions can be localized in phase space. This leads to the notion of wavefront sets, which refines the notion of singular support. 

\subsection{The Wavefront Set}
The wavefront set of a distribution describes simultaneously the location and `direction' of its singularities.
The starting point is the well-known fact that a compactly supported function or distribution is infinitely differentiable if and only if its Fourier transform decays as $O\bigl(\vert \xi \vert^{-m}\bigr)$ as $\vert \xi \vert \to \infty$ for every $m=1,2,\ldots$.
This characterizes the singular support $\ssupp(f) \subset \Real^n$, which is defined as the complement of the largest open set where $f$ is $C^{\infty}$. 

The singular support is however not invariant under smooth change of coordinates, so it is difficult to use it for understanding how an operator transforms the singularities of a distribution.
To address this drawback, we first localize the above characterization to a point $x_0 \in \Real^n$ by multiplying with a smooth cut-off that is non-zero at $x_0$.
A further localization (micro-localization) is obtained by identifying those `directions' in the frequency space where the already localized Fourier transform does not decay sufficiently fast. This singles out those directions (if there are any) that causes $f$ to be singular at $x_0$.
In particular, $f \in \mathcal{D}'(\Omega)$ has a singularity at $x_0 \in \Omega$ in direction $\xi_0 \in \Real^n \setminus \{ 0 \}$ if for any smooth cutoff function $\psi$ at $x_0$, the Fourier transform $\widehat{\psi f}$ does not decay rapidly in any open conic neighbourhood of the ray $\{ s \xi_0 : s > 0 \}$. 
The wavefront set of $f \in \mathcal{D}'(\Omega)$ can now be defined as the set of all tuples $(x_0,\xi_0)$ such that $f$ is singular at $x_0$ in direction $\xi_0$. 
With this definition, the wavefront set is a closed subset of $\Real^n \times (\Real^n \setminus \{ 0 \})$ that is conic in the second variable, i.e., it can be seen as a subset of $\Real^n \times S^{n-1}$.

The wavefront set can also be defined in an equivalent manner as a subset of the cotangent bundle of $\Real^n$ with the zero section removed, i.e., as a subset of $T^*(\Real^n)\setminus \{ 0\}$.
Such a definition has the advantage that it extends readily to distributions on smooth manifolds, which is required in many applications such as tomographic imaging.
We will therefore adopt the viewpoint where the wavefront set is a subset of the cotangent bundle. Hence, before stating the formal definition, we make a small digression into differential geometry.

Let $M$ be a smooth manifold and let $O_x$ denote the algebra of smooth functions defined in a neighbourhood of $x \in M$. 
Any functional $t \colon O_x \to \Real$ such that $t(f h)=t(f) f(x) + f(x) t(h)$ for $f,h \in O_x$ is called a tangent vector in $M$ at $x$.
The vector space of all tangent vectors at $x$, which is denoted by $T_x(M)$, is called the tangent space of $M$ at $x$. 
Its dual vector space, denoted by $T^*_x(M)$, is called the cotangent space of $M$ at $x$. 
An element $\tau$ of the cotangent space is called a covector (or differential form of degree 1) at $x$. 
Next, if $M_0 \subset M$ is sub-manifold of $M$, then $t$ is tangent to $M_0$ if $t(f) = 0$ for any function $f \in O_x$ that vanishes in $M_0$. Any functional $\tau \in T^*_x(M)$ such that $\tau(t) = 0$ for any tangent vector $t$ to $M_0$ is called a conormal covector to $M_0$ at $x$. 
Finally, the tangent bundle $T(M)$ and the cotangent bundle $T^*(M)$ are defined as the union of the tangent and cotangent spaces as $x \in M$ varies, i.e.,  
\[ T(M) := \bigcup_{x \in M} T_x(M) 
   \quad\text{and}\quad
   T^*(M) := \bigcup_{x \in M} T^*_x(M) = \bigl\{ (x,\eta) :  x \in M, \eta \in T_x^*(M) \bigr\}.
\]
\begin{definition}\label{def:WF:Manifold}
Let $M$ be a smooth manifold and $f \in \mathcal{D}'(M)$. We say that $f$ is \emph{microlocally smooth} at $(x_0,\xi_0) \in T^*(M)\setminus \{ 0 \}$ if there exists a neighbourhood $U \subset V$ of $x_0$ 
 and $\psi \in C^{\infty}_0(U)$ with $\psi(x_0) \neq 0$ and a conic neighbourhood $\Gamma$ of $\xi_0$ such that for constants $C_m$ we have 
 \begin{equation}\label{eq:WF:Smooth}
   \bigl\vert (\widehat{\psi f})(\xi) \bigr\vert
     \leq C_m \bigl( 1+ \vert \xi \vert\bigr)^{-m}
   \quad\text{for $\xi \in \Gamma$ with $m =1,2,\ldots$.}
 \end{equation}
 The \emph{$C^{\infty}$-wavefront set} $\WF(f) \subset T^*(M)\setminus \{ 0 \}$ is the set of $(x_0,\xi_0)$ where $f$ is not microlocally smooth.
\end{definition}
The wavefront set is a closed set that is conic in the $\xi$-variable, i.e., $(x,\xi) \in \WF(f)$ if and only if $(x,\lambda \xi) \in \WF(f)$ for any $\lambda > 0$. 
Next, it is clear that $f$ is equal to a $C^{\infty}$-function near $x$ whenever $(x,\xi)\not\in \WF(f)$ for all $\xi \neq 0$.
In fact, the $x$-projection of $\WF(f)$ equals the singular support of $f$ \cite[Proposition~8.1.3]{Hormander:2003aa}, so the wavefront set contains the information about the location of singularities.
Furthermore, $\WF(f)$ is also invariant under a diffeomorphic change of variables, which is not true for the singular support.

When $M \subset \Real^n$ is an open set, then the differentials $\differential x_1, \differential x_2, \ldots, \differential x_n$ are a basis of $T_x^*(\Real^n)$ for any $x \in \Real^n$.
Hence, any element in $T_x^*(\Real^n)$ can be written as 
\[  \xi \differential x = \xi_1 \differential x_1 + \ldots + \xi_n \differential x_n \in T_x^*(\Real^n) \quad\text{with $\xi \in \Real^n$.} \]
The tangent bundle of $M$ is therefore isomorphic to $M \times \Real^n$, so the cotangent bundle has the structure $T^*(M) = M \times (\Real^n)^*$ where $(\Real^n)^*$ is the space dual to $\Real^n$.

To give some examples, if $f \in \mathcal{D}'(\Real^n)$ is a smooth density on the hypersurface $x_n = 0$, i.e., $f(x) = f(x',x_n) = g(x')\delta_0(x_n)$ for some $g \in C^{\infty}(\Real^{n-1})$, then 
\[
   \WF(f) = \{ (x',0; 0,\xi_n) : x' \in \supp(g) \text{ and } \xi_n \neq 0 \}.
\]
Another example is when $f$ is the characteristic function of a domain $\Omega \subset \Real^n$ with smooth boundary. 
Then $\WF(f)$ is 
\[
 \WF(f) 
   = \bigl\{ (x,\xi \differential x) :  
       \text{$x \in \partial \Omega$ and $\xi$ is normal to $\partial \Omega$ at $x$} 
     \bigr\}.
\]
Furthermore, if $f$ is a linear combination of characteristic functions on sets with smooth boundaries, then $\WF(f)$ is the union of the conormal bundles to the individual sets unless cancellation occurs along shared boundaries. 
Further examples are given in \cite[Section~4]{Krishnan:2015aa}.

\subsection{Characterization of Visible Singularities}\label{subsec:MicroLocFIO}
Having defined the notion of a wavefront set, we now turn our attention to characterizing how an operator transforms the wavefront set.
This is one of the central questions in microlocal analysis and it turns out that useful results can be obtained for certain classes of operators. 

Let $M$ and $N$ be smooth manifolds and consider an operator $P \colon \mathcal{D}'(M) \to \mathcal{D}'(N)$. 
If $P$ is a differential operator, then it one can show that it does not increase the wavefront set. 
This is a special case of a more general result that shows the same result when $P$ is a pseudodifferential operator \cite[Theorem~14]{Krishnan:2015aa}:
\[
   \ssupp( P(f) ) \subset \ssupp(f)  \quad\text{and}\quad \WF\bigl( P(f) \bigr) \subset \WF(f)
\]
with equality if $P$ is in addition elliptic, i.e.,
 \[
   \ssupp( P(f) ) = \ssupp(f)  \quad\text{and}\quad \WF\bigl( P(f) \bigr) = \WF(f).
 \]  
Hence, a pseudodifferential operator may spread the support of a function $f$, but it does \emph{not spread} its singular support or \emph{wavefront set}. This is referred to as the \emph{pseudolocal property} and it has important consequences for inverse problems. In an inverse problem one often seeks to (approximately) invert a pseudodifferential operator $P$ by another operator $Q$ with an integral representation that is similar to $P$. The pseudolocal property would then imply that the singularities of $Q P f$ are identical to those of $f$, so this approximate inversion can be designed to recover the singularities of $f$.

Fourier integral operators (FIOs) is an even wider class of operators that contain the pseudodifferential operators as a special case.
FIOs are linear and defined in terms of amplitudes and phase functions, see \cite[Definition~7]{Krishnan:2015aa} for the precise definition. 
They arise naturally when one seeks to represent solutions of PDEs and in inverse problems, e.g., FIOs model the solution operators of hyperbolic PDEs as well as a variety of integral transforms, like the Radon transform and its generalizations. 
One of the main results of microlocal analysis is the H\"ormander-Sato Lemma \cite[Theorem~5.4]{Hormander:2003aa}, which states that if $P \colon \mathcal{D}'(M) \to \mathcal{D}'(N)$ is a FIO, then 
\begin{equation}\label{eq:HormanderSatoLemma}
  \WF\bigl(P(f)\bigr) \subset C \circ \WF(f) \quad\text{for $f \in \mathcal{E}'(M)$.}
\end{equation}
Here, $C \subset T^*(N) \times T^*(M)$ is the (microlocal) canonical relation associated to $P$, see \cite[eq.~(52)]{Krishnan:2015aa} for its formal definition and \cite[Definition~8]{Krishnan:2015aa} for how to interpret $C \circ D$ for $D \subset T^*(M)$. 
In \eqref{eq:RadonCanonical} we write out the canonical relation for the Radon transform, which is an example of a FIO that is not a pseudodifferential operator. 

\subsection{Applications to Tomographic Imaging}
\label{subsec:MicroLocCT}
Data in planar X-ray transmission tomography can in a simplified setting be modelled as samples of the (classical) 2D Radon transform. 
This is an integral transform that maps a function/distribution $f \in \mathcal{D}'(\Real^2)$ to a function/distribution $g$ on lines in $\Real^2$.
A more explicit definition requires introducing coordinates on the manifold of lines in $\Real^2$.
Any line $\ell \subset \Real^2$ can be described uniquely by $(\theta,p) \in [0,\pi] \times \Real$ through 
\[
  \ell = \bigl\{ x \in \Real^2 : x= s \mapsto p \omega(\theta) + s \omega(\theta)^{\bot} \bigr\}
  \quad\text{where}\quad
  \begin{cases} 
     \omega(\theta) := (\cos\theta, \sin\theta) \in S^1 & \\
     \omega(\theta)^{\bot} := (-\sin\theta,\cos\theta). &
  \end{cases}
\]
Hence, $(\theta,p) \in [0,\pi] \times \Real$ serves as coordinates for a sub-manifold $\datamanifold$ of lines in $\Real^2$, i.e., $\datamanifold \subset [0,\pi] \times \Real$.
Using these coordinates, we can express the Radon transform of $f \in \mathcal{D}'(\Real^2)$ as 
\[ 
  g = R(f)(\theta,p) := \int_{-\infty}^{\infty} 
    f\bigr(p \omega(\theta) + s \omega(\theta)^{\bot}\bigl) \,ds
  \quad\text{for $(\theta,p) \in \datamanifold$.}
\]
One can show that $R$ is a continuous map from $L^1(\Real^2)$ to $L^1(\datamanifold)$. 
There is a rich mathematical theory that investigates properties of such transforms. 
Particular focus is on injectivity, deriving explicit inversion formulas, and characterizing the range, see \cite{Markoe:2006aa} for a nice survey. 

A natural task in tomographic imaging is to reconstruct $f$ from $g=R(f)$, preferably by using a stable recovery scheme.
A closely related task is to recover the wavefront set of $f$ from the wavefront set of data $g\in \mathcal{D}'(\datamanifold)$, which is a subset of $T^*(\datamanifold)$.
This involves taking the Fourier transform of a localized version of $g$. 
To do that, we make use of the aforementioned coordinates on $\datamanifold$. 
One can extend $g$ periodically in $\theta$ and take localizing functions $\psi$ with support in $\theta$ less than a period, which allows us to view $\psi g$ as a function on $\Real^2$. 
Hence, its two-dimensional Fourier transform can be calculated using these coordinates. 
Next, let $d\theta$ and $dp$ be the standard basis of $T_{(\theta,p)}^*(\datamanifold)$ where the basis covector $d\theta$ is the dual covector to $\partial/\partial\theta$ and $dp$ is the dual covector to $\partial/\partial p$. 
One can also view $(x; r \differential x)$ as the vector $(x;r)$ where $x \in \Omega$ and $r$ is a tangent vector at $x$. In a similar way, $(\theta, s; a \differential\theta + b \differential p)$ can be viewed as the vector $(\theta, p; a, b)$.

The next theorem characterises those singularities of $f\in \mathcal{D}'(\Real^2)$ that can be recovered from Randon transform data $g=R(f)$, see \cite[Corollary~1]{Krishnan:2015aa} for its proof.
\begin{theorem}\label{thm:RadonWF}
Let $f \in \mathcal{D}'(\Real^2)$ and $g =R(f)$. 
Then
\[ 
  (x_0,\xi_0 \differential x) \in \WF(f)
  \quad\implies\quad
  \Bigl(\theta_0, x_0 \cdot \omega(\theta_0) ; s \bigl(-x_0 \cdot \omega(\theta_0)^{\bot}\differential\theta + \differential p \bigr) \Bigr) 
   \in \WF(g)
\]
whenever $\theta_0 \in [0,\pi]$ and $s \neq 0$ are such that $\xi_0 = s \omega(\theta_0)$.
Likewise, if $(\theta_0,p_0) \in [0,2\pi] \times \Real$ and $q \in \Real$, then 
\[  \bigl(\theta_0, p_0 ; s(-q \differential\theta+\differential p) \bigr) \in \WF(g)
    \quad\implies\quad
    (x_0,\xi_0 \differential x) \in \WF(f)
\]    
whenever $x_0 := p_0 \omega(\theta_0) + q \omega(\theta_0)^{\bot}$ and $\xi_0 := s \omega(\theta_0)$.
\end{theorem}
A key step in the proof is to show that the Radon transform is an elliptic Fourier integral operator.
One can then refer to a stronger version of the H\"ormander-Sato Lemma in \eqref{eq:HormanderSatoLemma} for elliptic FIOs \cite{Treves:1980aa} and conclude that 
\begin{equation}\label{eq:HormanderSatoLemmaRadon}
  \WF\bigl(R(f)\bigr) = C \circ \WF(f)
  \quad\text{whenever $f \in \mathcal{D}'(\Real^2)$.}
\end{equation}
Here, $C$ is the associated canonical relation and the next step is to explicitly calculate it for the Radon transform \cite[Theorem~17]{Krishnan:2015aa}:
\begin{multline}\label{eq:RadonCanonical}
 C = \Bigl\{ 
    \bigl( \theta, p, 
           s (-x \cdot \omega(\theta)^\bot \differential\theta+\differential p);
           x, s\omega(\theta) \differential x 
    \bigr) \in T^*(\datamanifold): 
    \\
    (\theta,p) \in \datamanifold,
      x \in \Real^2, s \neq 0, x \cdot \omega(\theta) = p
  \Bigr\}.
\end{multline}
The claims in Theorem~\ref{thm:RadonWF} now follows from combining \eqref{eq:HormanderSatoLemmaRadon} and \eqref{eq:RadonCanonical}.

Theorem~\ref{thm:RadonWF} implies in particular that the Radon transform $R$ detects singularities of $f$ perpendicular to a line included in the manifold $\datamanifold$. These singularities a referred to as visible, whereas singularities of $f$ in other directions do not show up in data $g=R(f)$ (invisible singularities).
In particular, if $f$ is a sum of characteristic functions of sets with smooth boundaries, then the tangent line to any point on the boundary of a region is normal to the wavefront direction of $f$ at that point. Hence, a singularity of $f$ at a boundary point $x$ is detectable if there is a line in $\datamanifold$ that is tangent to $x$. 

The above analysis for the Radon transform can also be applied to other operators relevant for tomographic reconstruction.
One is the backprojection $R^*$ which maps data to image space by integrating $g \in L^1(\datamanifold)$ over all lines in $\datamanifold$ that goes through $x$: 
\[
  R^*(g)(x) := \int_{0}^{2 \pi} g\bigl(\theta,x \cdot \omega(\theta)\bigr)\,d\theta
    \quad\text{for $g\in L^1(\datamanifold)$.}
\]
Clearly, $R^*$ maps data to an image but it is not an inverse or $R$.
However, applying $R^*$ to data $g = R(f)$ is the same as $R^* R(f)$, and it turns out that $R^* R$ is an elliptic pseudodifferential operator of order $-1$ \cite[Theorem~13 and eq. (42)]{Krishnan:2015aa}. 
Hence, by the general theory of microlocal analysis, the backprojection may spread out the support of the function $f$, but it does not spread out its singular support or wavefront set. 

Another related operator that maps the data to an image is the filtered backprojection (FBP) operator:
\[
  \FBP(g) := R^*( k \circledast g) 
  \quad\text{where $k = R^*(K)$.}
\]
The `$\circledast$' above is a 1-dimensional convolution along $p$-variable and if $g=Rf$, then $\FBP(g) = f \ast K$ where `$\ast$' is the 2-dimensional convolution. 
The idea in FBP is now to choose the reconstruction kernel $k$ such that $K \approx \delta$.
The FBP operator is, just like the backprojection, a pseudodifferential operator \cite[Example~8]{Krishnan:2015aa}, so it could spread the support of the function $f$, but it does not spread its singular support or wavefront set. 

We conclude with mentioning the lambda-reconstruction operator \cite[eq. (24)]{Krishnan:2015aa}. 
This operator is similar to FBP, but instead of recovering $f$ it aims to to recovers its singular support and wavefront set.
The main advantage of the lambda-reconstruction operator is that it is local, so one only needs to include lines through a neighbourhood of a point in order to recover the behaviour of $f$ at that point. 
The lambda-reconstruction operator is an elliptic pseudodifferential operator of order one, so it recovers the singular support and wavefront set $f$.

\section{Shearlets and Wavefront Sets}
\label{sec:ShearWFsets}

Due to a lack of a directional component, a (continuous) wavelet system is isotropic and hence not able to resolve the wavefront set of a distribution, in the sense of detecting it by decay properties of the wavelet coefficients of the distribution. Shearlet systems were introduced in as an anisotropic representation system with the structure of an affine system, consisting of a scaling operator to change the resolution, a translation operator to change the position, and a shearing operator to change the orientation, applied to a ``mother shearlet''. This construction indeed allows precise resolution of wavefront sets as we will discuss (see also \cite{kutyniok2009resolution}). 

\subsection{Continuous Case}
\label{subsec:ShearWFsetsCont}

For the definition of a continuous shearlet system, let $A_a$, $a \in \mathbb{R}^* := \mathbb{R} \setminus \{0\}$ be a {\em parabolic scaling matrix} $A_a$, $a \in \mathbb{R}^* := \mathbb{R} \setminus \{0\}$ and $S_s$, $s \in \mathbb{R}$ be a {\em shearing matrix} given by
\[
A_a = \begin{pmatrix} a & 0 \\ 0 & |a|^{1/2} \end{pmatrix}
\quad \mbox{and} \quad
S_s = \begin{pmatrix} 1 & s \\ 0 & 1\end{pmatrix}.
\]
Moreover, for $M \in \mathbb{R}^{2 \times 2}$, let the {\em dilation operator} be defined by
\[
D_M : L^2(\mathbb{R}^2) \to L^2(\mathbb{R}^2), \quad (D_M f)(x)  \mapsto |\det(M)|^{-1/2}f(M^{-1} x).
\]
Choosing $M$ in the dilation operator as $A_a$ and $S_s$ yields the set of scaling and shearing operators, respectively. Let us also emphasize that the choice of shearing instead of rotation is key to allowing a faithful digitalization due to the fact that the discrete versions $S_k$, $k \in \mathbb{Z}$, leave the digital grid $\mathbb{Z}^2$ invariant. Finally, let $T_t$, $t \in \mathbb{R}^2$, denote the 
{\em translation operator} as defined by
\[
T_t : L^2(\mathbb{R}^2) \to L^2(\mathbb{R}^2), \quad (T_t f)(x)  \mapsto f(x - t).
\]
This now leads to the following definition of continuous shearlet systems.

\begin{definition} \label{eq:CSS}
For $\psi \in L^2(\mathbb{R}^2)$, the {\em continuous shearlet system} $\mathcal{S}\mathcal{H}(\psi)$ is defined by
\[
\mathcal{S}\mathcal{H}(\psi) = \{ \psi_{a,s,t} := T_tD_{S_s}D_{A_a}  \psi = a^{-\frac34} \psi(A_a^{-1} S_s^{-1} (\, \cdot \,\, - t)):  a \in \mathbb{R}^*, s \in \mathbb{R}, t \in \mathbb{R}^2\}.
\]
\end{definition}

Group representation theory leads to conditions under which the associated shearlet transform is even an isometry. For this, let $\mathbb{S} := \mathbb{R}^* \times \mathbb{R} \times \mathbb{R}^2$ be endowed with the
group operation
\[
(a,s,t) \circ (a',s',t') = (a a', s+\sqrt{|a|}s', t + S_s A_a t').
\]
This is a locally compact group with left Haar measure $d_\mu(a,s,t) = da/|a|^3 ds dt$. 

\begin{theorem}[\cite{kutyniok2009coorbit}]
Let $\psi \in L^2(\mathbb{R}^2)$ be {\em admissible}, i.e., it satisfies
\[
\int_{\mathbb{R}^2} \frac{|\hat{\psi}(\xi)|^2}{|\xi_1|^2} \, d  \xi  < \infty.
\]
Then the {\em continuous shearlet transform} $SH_{\psi}: L^2(\mathbb{R}^2) \rightarrow L^2(\mathbb{S})$  given by
\[
SH_{\psi} f(a,s,t) = \langle f,\psi_{a,s,t} \psi \rangle.
\]
is an isometry.
\end{theorem}

One example for a suitable function $\psi \in L^2(\mathbb{R}^2)$ are {\em classical shearlets}, which are defined by
\[
\hat{\psi}(\xi) = \hat{\psi}(\xi_1,\xi_2) = \hat{\psi}_1(\xi_1) \, \hat{\psi}_2(\tfrac{\xi_2}{\xi_1}).
\]
Here, $\psi_1 \in L^2(\mathbb{R})$ is a wavelet, i.e., it satisfies the discrete Calder\'{o}n condition given by
\[
\sum_{j \in \mathbb{Z}}|\hat\psi_1(2^{-j}\xi)|^2 = 1 \quad \mbox{for a.e. } \xi \in \mathbb{R},
\]
with $\hat{\psi}_1 \in C^\infty(\mathbb{R})$ and  $\supp \, \hat{\psi}_1 \subseteq [-\frac54,-\frac14] \cup [\frac14,\frac54]$,
and $\psi_2 \in L^2(\mathbb{R})$ is a `bump function', namely
\[
\sum_{k = -1}^{1}|\hat\psi_2(\xi+k)|^2 = 1 \quad \mbox{for a.e. } \xi \in [-1,1],
\]
satisfying $\hat{\psi}_2 \in C^\infty(\mathbb{R})$ and $\supp \, \hat{\psi}_2 \subseteq [-1,1]$. 

\begin{figure}[!ht]
\centering
\includegraphics[width = 0.38\textwidth]{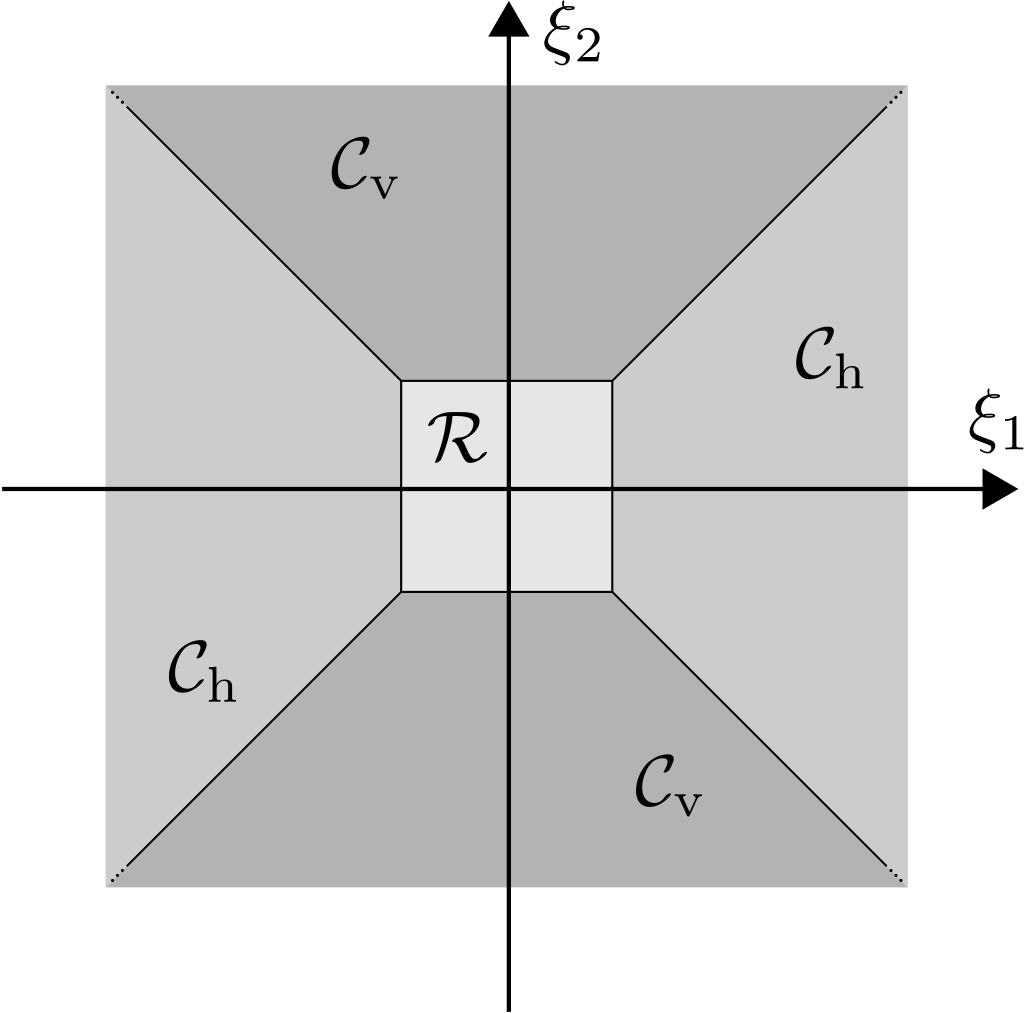}
\caption{Frequency cones for the cone-adapted shearlet system.}
\label{fig:Cones}
\end{figure}

Since the just defined continuous shearlet system exhibits a directional bias and thus is in this pure form not able to resolve any wavefront set, we require s slightly adapted version. This is based on a suitable splitting of the frequency domain into four conic regions and a low frequency part as illustrated in Figure~\ref{fig:Cones}.  This leads to the so-called cone-adapted continuous shearlet system. 

\begin{definition} 
For $\phi, \psi, \tilde{\psi} \in L^2(\mathbb{R}^2)$, the {\em cone-adapted continuous shearlet system}
is defined by $\mathcal{S}\mathcal{H}(\phi,\psi,\tilde{\psi}) = \Phi(\phi) \cup \Psi(\psi) \cup \tilde{\Psi}(\tilde{\psi})$,
where
\begin{eqnarray*}
\Phi(\phi) &=& \{\phi_t  := \phi(\cdot-t) : t \in \mathbb{R}^2\},\\
\Psi(\psi) &=& \{\psi_{a,s,t}=  a^{-\frac34} \psi(A_a^{-1}S_s^{-1}(\, \cdot \, -t))
: a \in (0,1],\, |s| \le  1+a^{1/2},\, t \in \mathbb{R}^2\},\\
\tilde{\Psi}(\tilde{\psi}) &=& \{\tilde{\psi}_{a,s,t}:=  a^{-\frac34} \tilde{\psi}(\tilde{A}_a^{-1}S_s^{-T}(\, \cdot \, -t))
: a \in (0,1],\, |s| \le  1+a^{1/2},\, t \in \mathbb{R}^2\},
\end{eqnarray*}
and $ \tilde{A}_a = $ diag$(a^{1/2}, a)$.
\end{definition}

Similar to continuous shearlet systems as defined in Definition \ref{eq:CSS}, also for cone-adapted continuous shearlet systems an associated transform can be defined, namely
$$
SH_{\psi} f(a,s,t) := \langle f,\psi_{a,s,t} \psi \rangle \quad \mbox{and} \quad
SH_{\tilde{\psi}} f(a,s,t) := \langle f,\tilde{\psi}_{a,s,t} \psi \rangle,
$$
and isometry conditions can be proven \cite{kutyniok2009resolution}. With this system, we are also now able to precisely resolve a wavefront set of a distribution in the following way.

\begin{theorem}[Theorem 5.1. of \cite{kutyniok2009resolution}]
Let $\psi \in L^2(\mathbb{R}^2)$ be admissible, and $f \in L^2(\mathbb{R}^2)$. Let $\mathcal{D} = \mathcal{D}_1 \cup \mathcal{D}_2$, where $\mathcal{D}_1 =\{(t_0,s_0) \in \mathbb{R}^2 \times [-1,1] :$ for $(s,t)$ in a neighborhood $U$ of $(s_0,t_0)$, $|SH_{\psi} f(a,s,t)| =O(a^k)$ as $a \to 0$, for all $k \in \mathbb{N}$, with the $O(\cdot)$--term uniform over $(s,t) \in U\}$ and $\mathcal{D}_2=\{(t_0, s_0) \in \mathbb{R}^2 \times [1,\infty) :$ for $(\frac 1s,t)$ in a neighborhood $U$ of $(s_0,t_0)$, $|SH_{\tilde{\psi}} f(a,s,t)| =O(a^k)$ as $a \to 0$, for all $k \in \mathbb{N}$, with the $O(\cdot)$--term uniform over $(\frac 1s,t) \in U$\}. Then
\[ \mbox{WF}(f)^c = \mathcal{D}.\]
\end{theorem}

\subsection{Digital Case as a Semantic Edge Detection Problem}
\label{subsec:ShearWFsetsDig}

Since we are now familiar with the approach to extract the wavefront set of a continuous distribution by analyzing the asymptotic behavior of the shearlet coefficients at a fixed position-orientation pair, we now aim to extend it to real world data. To be precise, we will now consider data coming from a finitely sampled function such as images formed by pixels representing point samples of a real-valued function. 

In the most general case, we showed in \cite{andrade2019wfset} that the previously discussed approach is not directly transferable to the situation of functions being defined on a grid on a bounded domain. Coarsely speaking, the reason for this is the fact that we just have access to a finite number of Fourier samples as well as finite number of shearlet coefficients. 

In order to overcome this limitation, we are now assuming that a digital image arises from the finite sampling of a continuous model in that sense that the image has itself a wavefront set in the sampling limit. As in our paper \cite{andrade2019wfset}, we aim to approximate the wavefront set by a sequence of what we call \emph{the digital wavefront sets}. 

Let us now define what we mean by a digital wavefront set (see also \cite{andrade2019wfset}). In the sense of Shannon's sampling theorem one can make use of Paley-Wiener spaces to define a sampling space of $L^2(\mathbb{R}^2)$ for the coarsest scale $\Lambda > 0$, namely $\PW_{\Lambda} \subset L^2(\mathbb{R}^2)$ defined by
\[
  \PW_{\Lambda} \coloneqq
  \left\{ f \in L^2\left(\Real^2\right) \colon \supp\big(\widehat{f} \, \big) \subset [-\Lambda,\Lambda]^d \right\}.
\]
Using this definition of the sampled space, we now define the notion of a {\em digital wavefront set extractor} for the given coarsest scale $\Lambda>0$ to be the map
\begin{align}\label{eq:OneShotExtractor}
\DWF_\Lambda: \PW_\Lambda \to P\left(\Real^2 \times \mathbb{S}^1 \right) \quad\text{such that} \quad \DWF(P_{\Lambda} f ) = \WF(f) \mbox{ for all } f \in L^2(\Real^2).
\end{align}

The existence of a so-called {\em faithful sequence of digital wavefront set extractors} assumes that the sequence of maps $\{\DWF_j\}_{j\in \mathbb{N}}$ converges to the continuous wavefront set extractor $WF$ in the Hausdorff sense, i.e., we have
\begin{align}\label{eq:clairvoyance}
d_H\left( \overline{\DWF_j( P_j(f) )_x}, \overline{\WF(f)_x} \right) \to 0.
\end{align}
In \cite{andrade2019wfset}, we make use of Shannon's sampling theorem to show that it is not possible to analytically define a sequence of digital wavefront set extractors for general function classes, which are dense on $L^2(\mathbb{R}^2)$ functions (see \cite[Thm. 3.3]{andrade2019wfset}). In this work, we also indicated that, although for most classical function classes it is impossible to construct an analytical digital wavefront set extractor on their Paley-Wiener space projections, for function classes in applications such as natural images, which are typically empirically defined, such wavefront set extractor could exist. However, those will certainly be highly sensitive to the choice of the class.

Inspired by these results, it is natural to have as \emph{guiding principle} of the construction of such digital wavefront set extractors, in our case the wavefront set extractor should be closely adapted to the underlying function class. 

As mentioned, typical function classes arising from real-world applications are empirically defined. Thus most of model-based heuristics that could potentially be used to construct analytically a digital wavefront set extractor for each class, will fail with high probability.

Using solely, model-based heuristics will limit the model with rigid assumptions and the guiding principle wont be followed. Although the best choice for our guiding principle, will be to learn the data representation from scratch this approach will typically be computationally intractable, requiring a lot of data and complicated architecture, having as example, CASENet \cite{yu2017casenet} and SEAL \cite{yu2018simultaneous}. 

Summarizing, our approach presented in \cite{andrade2019wfset} follows the main thrust of current deep learning-based methodologies in imaging science. It leads to a state-of-the-art wavefront set extraction by combining the model-based digital shearlet transform as a pre-processor for edge detection and a convolutional neural network that locally learns the wavefront set by classifying patches of the digital shearlet coefficients with the potential singular point at the center. 

\subsubsection{Digital Shearlets}

Let us now delve more deeply into the digital shearlet transform \cite{kutyniok2012digital} for a digital domain of pixel images, which --- as explained before --- is used in the classifier proposed in \cite{andrade2019wfset}. The construction of the digital shearlets is inspired by the fast wavelet transform. In fact, the digital function system can be seen as a filter bank, for which the novelty in the shearlet case resides in the definition of a faithful digital shearing operator.

To explain the digital shearlet transform in detail, let $M\in \Natural$, $J \subset \Natural$ be finite, $k_j \subset \Natural$ for all $j \in J$ and $K_j \coloneqq [-k_j, \dots, 0, \dots, k_j]$. We then pick $2 \sum_{j \in J} K_J +1$ matrices in $\Real^{M \times M}$, and denote these matrices by $\phi^{dig}$ and $\psi_{j,k, \iota}^{dig} \text{ for } j \in J, k \in K_j, \iota \in \{ -1,1\}$. To make the connection to the continuous shearlet transform, we can think of $\psi_{j,k, \iota}^{dig}$ as a digitized version of $\psi_{2^{-j}, 2^{-j/2}k, 0, \iota}$ and of $\phi^{dig}$ as a digitized version of a low frequency filter.
An explicit --- and highly technical --- construction of the matrices $\phi^{dig}$ and $\psi_{j,k, \iota}^{dig}$ can be found in \cite{kutyniok2012digital}. Those are then exploited to define the \emph{digital shearlet transform} of an image $I \in \Real^{M \times M}$ by
\[
\mathrm{DSH}(I)(j,k,m,\iota) \coloneqq \begin{cases}
  \bigl\langle  
    I, T_m \psi_{j,k, \iota}^{dig}
  \bigr\rangle,
  & \text{ if $\iota \in \{-1,1\}$,} \\[0.5em]
  \bigl\langle  
    I, T_m \phi^{dig} 
  \bigr\rangle, 
  & \text{ if $\iota = 0$,}
\end{cases}
\]
where $j \in J, k \in K_j$, $m \in \{1, \dots, M\}^2$, and $T_m: \Real^{M \times M} \to \Real^{M \times M}$ circularly shifts the entries of the elements of a matrix by $m$. Thus, from a structural viewpoint, the digital shearlet transform of an image $I \in \Real^{M \times M}$ is a stack of $2 \sum_{j \in J} (K_j-1) +1$ matrices of dimension $M \times M$. We will refer to this stack of images, as the shearlet volume.

\subsubsection{DeNSE Algorithm}

Using the digital shearlet transform as the input of a convolutional neural network classifier, the Deep Network shearlet Edge Extractor (DeNSE) is able to perform wavefront set extraction with high accuracy and closely adapted to the class of the training data. In addition to follow the guiding principle, the main purpose of this approach was to achieve high accuracy concerning the task of wavefront set extraction for inverse problem regularization. With this in mind, the problem of classifying pixels of a digital image into $N+1$-classes, where $N$ of these classes correspond to orientations of singular points and the additional ($N+1$th) class is the binary class of a pixel being an edge, is splitted into $N+1$ separate binary classifiers. In addition, each pixel is classified independently as well by using a patch-based approach in the sense that a pixel will be classified to be an edge of a certain orientation by classifying a patch of the shearlet coefficients where the pixel is centered at.   

\begin{figure}[h!]
\centering
\begin{tikzpicture}[x={(1,0)},y={(0,1)},z={({cos(60)},{sin(60)})},
font=\sffamily\small,scale=2]
%

\foreach \X [count=\Y] in {1.5,1.5,1.3,1.1}
{\draw pic (box1-\Y) at (\Y*0.75,-\X/2,0) {fake box=white!70!gray with dimensions 0.35 and {2*\X} and 1*\X};
\node[draw,single arrow, black,fill=black!50] at (\Y*0.75+0.45,0.2,0) {\scriptsize{ReLU}};
}

\foreach \X/\Col in {6.3/red,6.5/red,6.7/green,6.9/green, 7.1/blue, 7.3/blue}
{\draw[canvas is yz plane at x = \X*0.7, transform shape, draw = red, fill =
\Col!50!white, opacity = 0.5] (0,0.5) rectangle (2,-1.5);}

\draw[gray!60,thick] (6.1*.7,-0.1,-1.6) coordinate (1-1) -- (6.1*.7,-0.1,0.6) coordinate (1-2) -- (6.1*.7,2.,0.6) coordinate (1-3) -- (6.1*.7,2.1,-1.6) coordinate (1-4) -- cycle;
\draw[gray!60,thick] (7.5*.7,-0.1,-1.6) coordinate (2-1) -- (7.5*.7,-0.1,0.6) coordinate (2-2) -- (7.5*.7,2.,0.6) coordinate (2-3) -- (7.5*.7,2.1,-1.6) coordinate (2-4) -- cycle;
\foreach \X in {4,1,3}
{\draw[gray!60,thick] (1-\X) -- (2-\X);}

\node[draw,single arrow, orange,fill=orange!30] at (7.7*0.7,0.5,0) {Flatten};

\node[circle,draw,blue,fill=blue!30] (A1) at (8.6*0.7,1,0) {~~~};
\node[circle,draw,red,fill=red!30,below=4pt of A1] (A2) {~~~};
\node[circle,draw,green,fill=green!30,below=18pt of A2] (A3) {~~~};
\draw[circle dotted, line width=2pt,shorten <=3pt] (A2) -- (A3);

\node[circle,draw,fill=gray!60] (B1) at (10*0.7,1,0) {~~~};
\node[circle,draw,fill=gray!60,below=4pt of B1] (B2) {~~~};
\node[circle,draw,fill=gray!60,below=18pt of B2] (B3) {~~~};
\draw[circle dotted, line width=2pt,shorten <=3pt] (B2) -- (B3);

\node[circle,draw,fill=gray!60] (C1) at (11*0.7,0.7,0) {~~~};
\node[circle,draw,gray,fill=gray!60,below=7pt of C1] (C2) {~~~};

\begin{scope}[on background layer]
\node[orange,thick,rounded corners,fill=orange!30,fit=(A1) (A3)]{};
\node[gray,thick,rounded corners,fill=gray!10,fit=(B1) (B3)]{};
\node[gray,thick,rounded corners,fill=gray!10,fit=(C1) (C2)]{};
\end{scope}

\foreach \X in {1,2,3}
{\draw[-latex] (A\X) -- (B1);
 \draw[-latex] (A\X) -- (B2);
 \draw[-latex] (A\X) -- (B3);
 \draw[-latex] (B\X) -- (C1);}
\end{tikzpicture}
\caption{Illustration of the network architecture forming the foundation of the classifier. This network consists of four convolutional layers and one fully-connected layer. The colored block in the middle represents a stack of the output of the last convolutional layer. The colors correspond to the different channels.}
\label{fig:NetworkArch}
\end{figure}
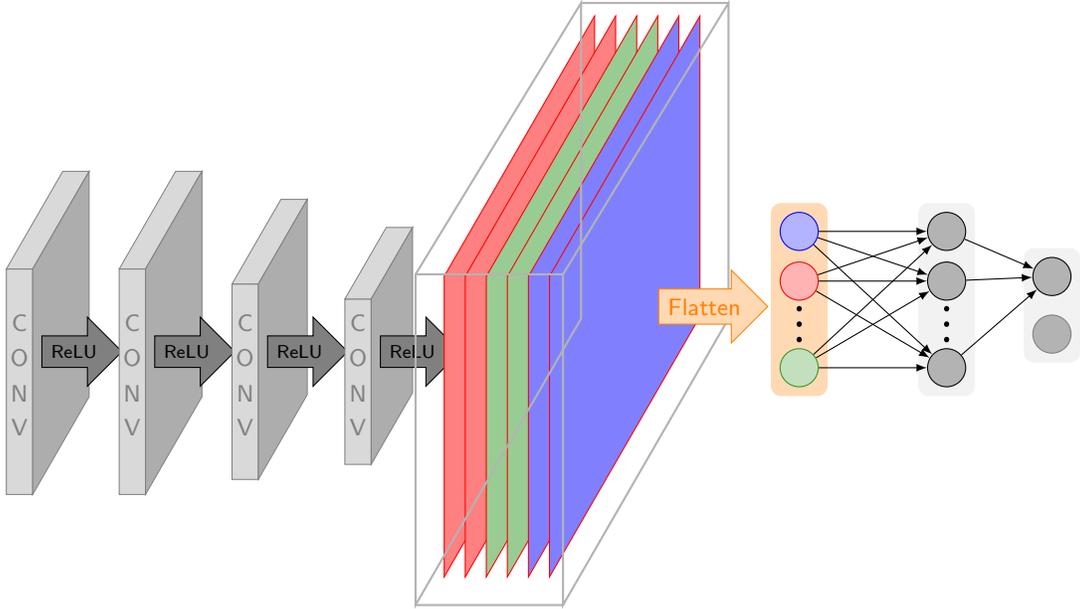

The architecture used for each classification was composed of four convolutional layers, with $2\times 2$ max pooling, ReLU activation, and batch normalization, followed by a fully connected layer with 1024 neurons, softmax activation function, and a one dimensional output. We chose this architecture
since it performed well in a series of tests while being of moderate size. The network architecture is illustrated in Figure~\ref{fig:NetworkArch}. 

As in \cite{andrade2019wfset}, if we pick $J=4$ as the coarsest scale, the digital shearlet coefficients $DSH(I)$ will form a three dimensional array composed by $49$ stacked digital images of the same size as the original image. Having these coefficients, if the patches to classify are as in the original work, of size $21\times 21$, the input of DeNSE will be a stacked set of patches of size $21\times 21\times 49$.

If for example, we have 180 orientations $\{\theta_i\}_{i=1}^{180}$, for each $\theta_i$, the network is trained on patches of shearlet coefficients of images $I\in \mathbb{R}^{M\times M}$ of the form
\begin{align}\label{eq:shearletBatches}
(\mathrm{DSH}(I)(j,k,m,\iota))_{j \in J, k \in K_j, \iota \in \{-1,0,1\}, m \in [m^*_1-10, m^*_1+10] \times [m^*_2-10, m^*_2+10]},
\end{align}
with $m^*\in\{ 11, \ldots, M-10\}^2$ being the center point of the patch to which the classified orientation is assigned. By using a softmax binary classifier, the label associated to a batch of~\eqref{eq:shearletBatches} is $1$ if $I$ has a singularity with orientation $\theta_i$ at $m^*$. In addition, there exists a separate classifier, which uses the same data and assigns $1$ to a patch whose central pixel corresponds to a singular point (edge point).

The DeNSE algorithm for wavefront set extraction \cite{andrade2019wfset} uses as coarsest scale $J = 4$ with the corresponding shearing levels for each scale given by $K_j = 2^{\lceil j/2 +1\rceil}+1$, obtaining digital shearlet coefficients arrange by $L$ stacked images, where $L = 2\sum_{j\in J} (K_j-1)+1 = 49$. The implementation of choice was the julia API of the software ShearLab \cite{Kutyniok:2016:SFD:2888419.2740960} (\url{www.shearlab.org/software}).

The Deep Network shearlet Edge Extractor (DeNSE) was in fact shown to outperform other methods, for example, CASENet and SEAL, on the standard datasets such as BSDS500 (Berkeley segmentation dataset) with 503 natural images, the semantic boundaries dataset (SBD) with 11355 natural images, and a set of phantoms formed by ellipses and paralellograms with analytically defined wavefront set. We present some of those results in Section~\ref{sec:NumResApp}. 

\subsection{Avoiding the Distracted Supervision Paradox}
\label{subsec:ShearWFsetsParad}

The distracted supervision paradox as introduced in Section~\ref{subsec:IntroDataDriveEdge} refers to the fact that semantic edge detection such as wavefront set extraction requires the supervision of two fundamentally different tasks:
\begin{itemize}
    \item \textbf{Category-agnostic edge detection} requires the detection of pixels corresponding to edges and other singularities, thus the use of low-level features.
    \item \textbf{Semantic edge classification} requires the classification of edges with abstracted high-level semantics, therefore relies on high-level features.
\end{itemize}
Intuitively accomplishing both tasks jointly seems infeasible as both problems require very different features. Yu et al. confirmed in \cite{yu2017casenet} with the CASENet architecture that a naive joint supervision of both tasks, performs less well than directly learning the semantic edges with no deep supervision that combines both features.

This paradox imposes an upper bound on the performance for methods that are aiming to directly learn (in an end-to-end fashion) semantic edges. Since the initial work \cite{yu2017casenet}, there have been several approaches to avoid this paradox. One particularly remarkable approach is the one presented by Liu et al. in \cite{yun2019sed}. In their paper, they propose a network architecture containing a backbone based on residual convolutional neural networks --- similar to the CASENet, but with the introduction of a novel information converter layers ---, which allows to combine information coming from lower levels used for edge supervision with information from higher levels used for semantic supervision, which is ultimately guided by the detected edges. This approach successfully established a new state-of-the-art in semantic edge detection, with a significant performance improvement. 

The key to the Deep Network Shearlet Edge Extractor (DeNSE) resides on the splitting of the multi-label classification task into individual binary classifiers inspired on the performance increment. In addition DeNSE separates the category-agnostic edge detection and the semantic edge classification, which already avoids the distracted supervision paradox.  

\section{General Semantic Edge Detection Using Shearlets and Deep Supervision}
\label{sec:Gensemantic edge detectionShearDS}

The use of shearlets for achieving high precision in digital wavefront set extraction motivates the introduction of the shearlet transform in general semantic edge detection, where classes of edges are coming typically from the particular object the edge belongs to. 
In order to illustrate the power of the shearlet transform as a preprocessing step, we are going to make use of the backbone of the current state-of-the-art approach on general semantic edge detection, based on the CASENet architecture \cite{yu2017casenet}. As already mentioned, after the introduction of CASENet and the distracted supervision paradox, most of the approaches to perform semantic edge detection, made use of the same architecture backbone. Each model, introduced an alternative way to train it, mostly to avoid the distracted supervision paradox.


At the following we explore two alternatives architectures, namely the original CASENet and the deep diverse supervision approach presented by Liu et al in \cite{yun2019sed}, where information converters are used for the joint training of low-level edge supervision and high-level edge classification. In order to show the heavy lifting that the shearlet transform is able to do, we use as the input for these networks the shearlet coefficients of the images, and extend the channels of the first convolutional layer by the number of slices of the particular shearlet coefficients. We also decrease the depth of the resulting network by removing the last category-agnostic edge feature map resnet subnetworks, resulting in an overall smaller network, which will be addressed in detail in the next sections. 

\section{Shear-CASENet: Deep Shearlet Category-Aware Semantic Edge Detection}

\begin{figure}[h!]
\centering
\includegraphics[width = 0.38\textwidth]{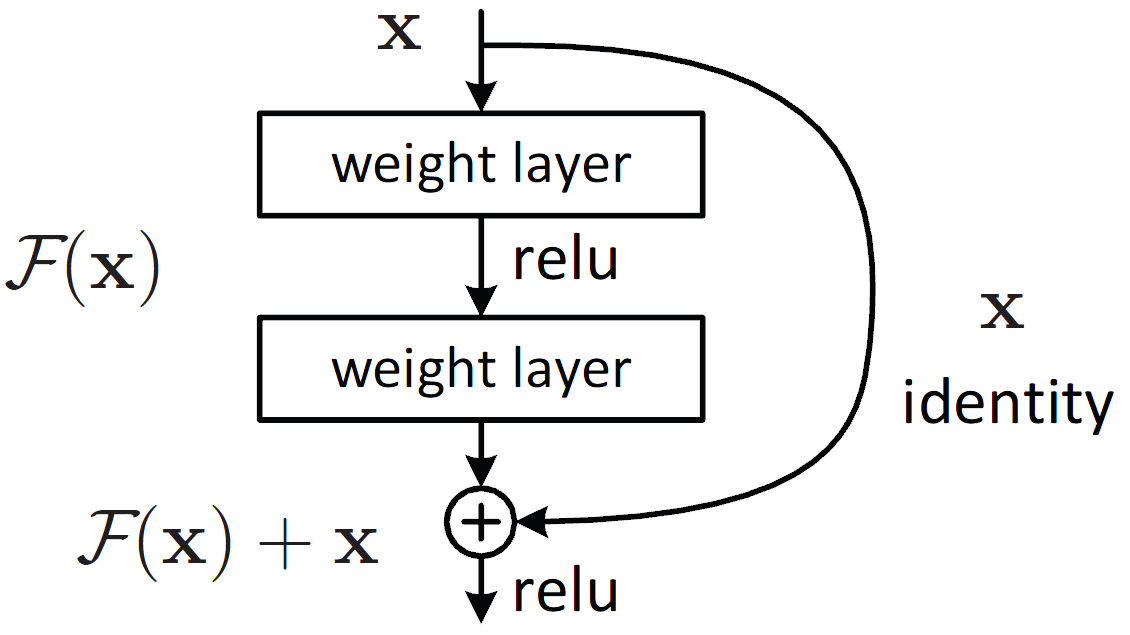}
\caption{Illustration of the principal block in ResNet, namely the skip connection from the input to the output is the main characteristic of this architecture (taken from \url{https://neurohive.io/en/}).}
\label{fig:ResNetBlock}
\end{figure}

The CASENet architecture \cite{yu2017casenet} is based on the already known Residual Neural Network architecture, known as ResNet (see Figure~\ref{fig:ResNetBlock}). This architecture has shown tremendous success in different image processing task, including image classification. It in fact did even won the ImageNet challenge in 2015. CASENet receives as input the image and it produces as output a two-dimensional array of the same size with the classified edges represented as pixels with value given by the corresponding category.

\begin{figure}[h!]
\centering
\includegraphics[width = 1\textwidth]{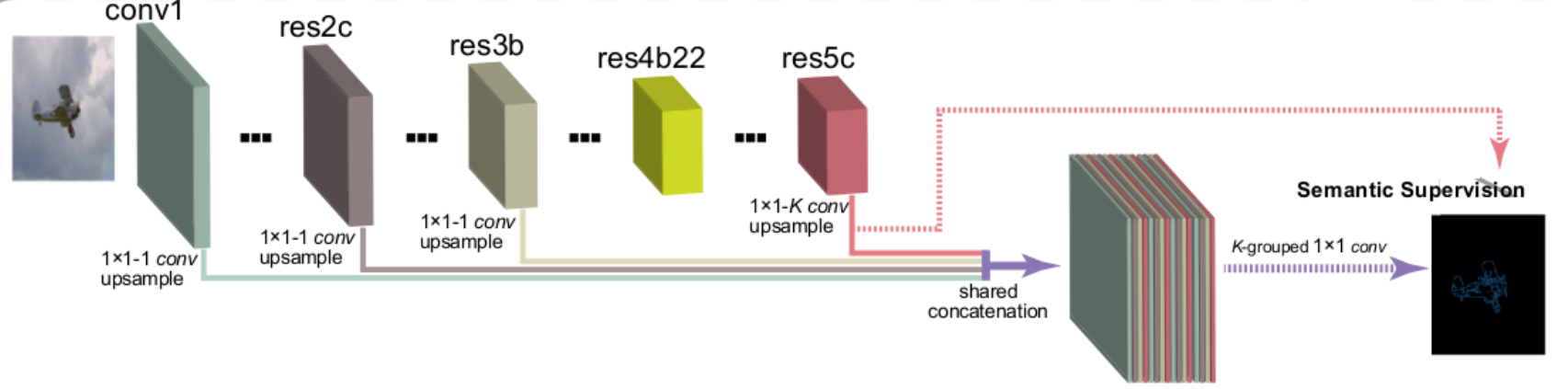}
\caption{Illustration of the classical CASENet architecture (inspired by the figures of \cite{yun2019sed}).}
\label{fig:CASENet}
\end{figure}

We next explain the CASENet architecture, which is displayed in Figure~\ref{fig:CASENet}, in more detail. The input image is connected to a $1-channel$ convolutional layer (conv1), which is followed by four stacked ResNet subnetworks; res2c, res3b, res4b22 and res5c correspondingly. Each of those sub-networks is a block of the network ResNet-101, where res(N)(M) represents the M-th layer (represented by the letter "a", "b" and "c") of the N-th stage of ResNet-101. 

The first three stages of CASENet (i.e. conv1, res2c, res3b) produce a single channel feature map $F^{(m)}$, which is used to perform the edge detection part. The last stage, res5c, is connected to a $1\times 1$ convolutional layer to produce a $K$-channel class activation map $A^{(5)} = \{A^{(5)}_1,A^{(5)}_2, \ldots, A^{(5)}_K\}$, where $K$ is the total number of categories. 
In order to combine the edge information coming from the first stages, at the end of the network, one replicates the bottom features $F^{(m)}$, by concatenating them in each channel of the class activation map at the last stage, namely:
$$
A^f = \{ F^{(1)}, F^{(2)}, F^{(3)}, A^{(5)}_1, \ldots, F^{(1)}, F^{(2)}, F^{(3)}, A^{(5)}_K\}.
$$
At the end, a $K$-grouped $1\times 1$ convolutional layer is applied to $A^f$, generating a semantic edge map with $K$ channels, whear the $k$-th channel represents the edge map for the $k$-th category. 
Summarizing, the first four stages of CASENet produce category-agnostic edge feature maps with different levels of refinement. This depth becomes necessary to produce edges fine enough to be classified by the last stage. 

Recall that he shearlet transform has shown good performance as a feature extractor for edge detection, due to its properties on wavefront set resolution. With this in mind, it seems conceivable that applying CASENet to the shearlet coefficients instead of the image itself, the produced edge in the first feature maps will be fine enough already in first stages.

\begin{figure}[h!]
\centering
\includegraphics[width = 1\textwidth]{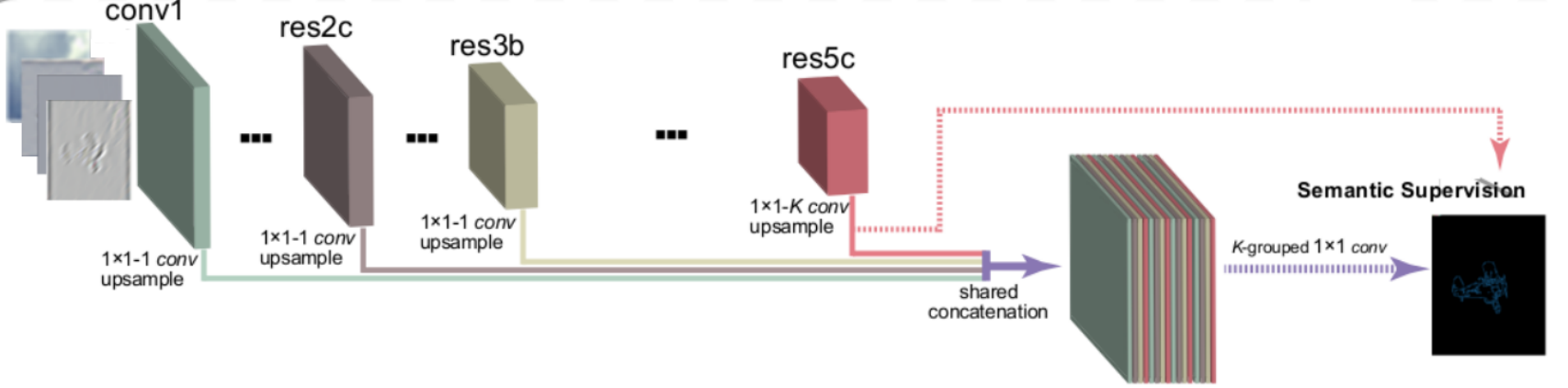}
\caption{Illustration of the Shear-CASENet architecture (inspired in the figures of \cite{yun2019sed}).}
\label{fig:ShearCaseNet}
\end{figure}

By now combining shearlets and the CASENet architecture, we introduce the Shear-CASENet architecture, which takes as input the shearlet coefficients of an image and produces as output an array with the same size as the image with the classified edges, just like CASENet. 
Shear-CASENet, uses the same backbone as CASENet, extending the first convolutional layer with the same number of channels as slices of the shearlet volume and removing the fourth stage. Figure~\ref{fig:ShearCaseNet} depicts the Shear-CASENet architecture.

Shear-CASENet  has significantly less parameters to train than the classical CASENet. Indeed, if the input is an $N\times N$-image, the fourth stage of CASENet will have $1024\times N^2/8$ parameters. By having that stage removed and extending the first convolutional layer by $J$ shearlet channels, one obtains a reduction of $1024\times N^2/8-J$ trainable parameters. Typically four scales in the shearlet transform are used, i.e., $J=49$, leading to an image of size $256\times 256$. Shear-CASENet has $37259387$ parameters while the classical CASENet has $42436731$, with around $13\%$ less parameters to train. 

In addition, Shear-CASENet presented also better performance in semantic edge detection on the Semantic Boundaries Dataset (\url{http://home.bharathh.info/pubs/codes/SBD/download.html}). We will present the corresponding numerical experiments in detail in Section~\ref{subsec:NumResAppsemantic edge detection}.

\section{Shear-DDS: Shearlet Diverse Deep Supervision}

In 2017 the CASENet architecture imposed by its novel approach a new state-of-the-art in semantic edge detection. The authors also introduced the concept of distracted supervision paradox, by noticing the fundamental limitation of training jointly, with deep supervision, the category-agnostic edges and the edge classification. In CASENet, even the fourth stage is not used in the supervision. It was in fact introduced as a way to alleviate the information conflicts coming from the first three stages and the fifth stage, also known as a buffer unit. 

After the introduction of CASENet, different alternatives to train the network have been introduced. Yu et al. \cite{yu2018simultaneous} introduced the Simulatenous Edge Alignment and Learning (SEAL) architecture, which is a new training approach for the CASENet architecture. It simultaneously aligns the ground truth edges and learns the corresponding classifier, with the downside of being time consuming due to the necessary CPU usage by the alignment step. 

\begin{figure}[h!]
\centering
\includegraphics[width = 1\textwidth]{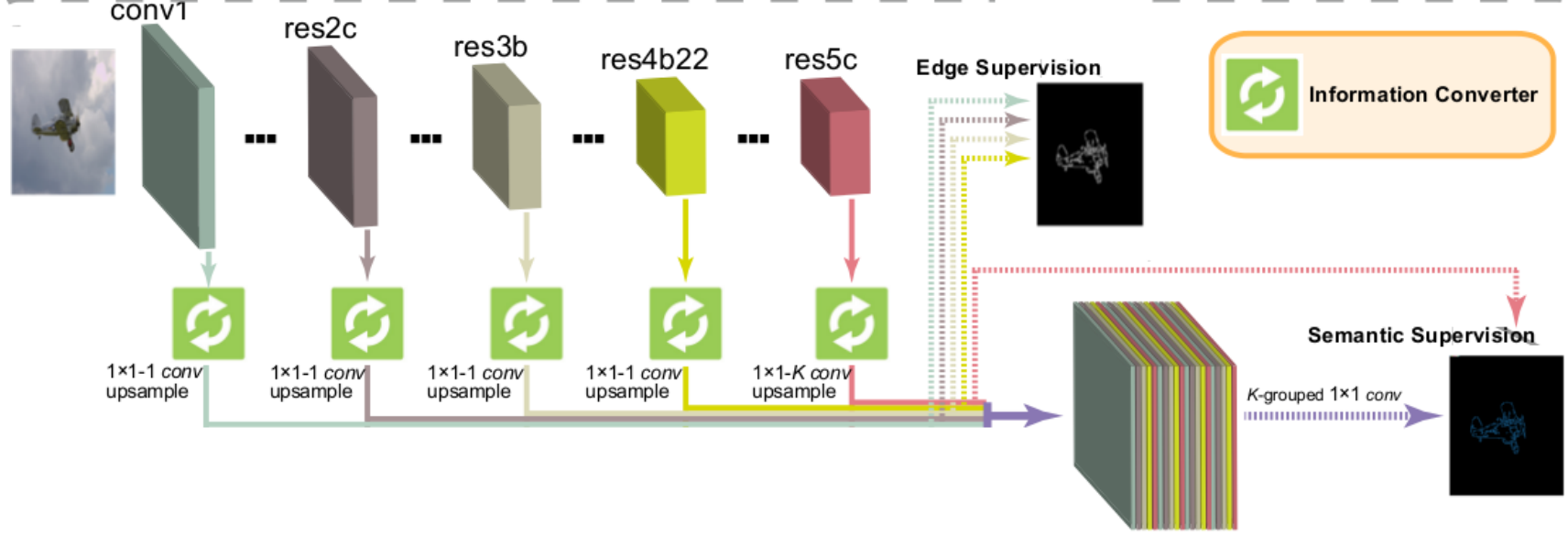}
\caption{Illustration of the classical Diverse Deep Supervision architecture (inspired by the figures of \cite{yun2019sed}).}
\label{fig:DDS}
\end{figure}

Recently Liu et al. introduced a novel way to train CASENet \cite{yun2019sed}, also known as the deep diverse supervision. This approach makes use of an information converter based on a convolutional residual block (see figure~\ref{fig:ResNetBlock}), where the output of each stage of CASENet is fused in a final shared concatenation. Figure~\ref{fig:DDS} depicts this architecture, it is worth to notice that in this case, stage four is not anymore a buffer, but it is already used in the supervision.

The information converters help to assist low-level feature learning (stage one-four) in order to generate consistent gradient signals from the higher levels (stage five), producing a highly discriminative feature map for high performance semantic edge detection. 

Having the category-agnostic edge maps obtained from the information converter applied to each of the first four stages, namely $E = \{ E^{(1)}, E^{(2)}, E^{(3)}, E^{(4)}\}$, the final map will be given by the information conversion of the fifth stage and the shared concatenation, i.e.,
$$
E^f = \{ E, A_1^{(5)}, E, A_2^{(5)}, \ldots E, A_K^{(5)}\}
$$

This network is trained with a multi-task loss, meaning, two different losses, corresponding to category-agnostic and category-aware edge detection, are optimized jointly. Both losses are based on reweighted sigmoid cross-entropy loss, which is typically used for multi-label classification. For further details, we refer to \cite{yun2019sed}.

Using a similar approach as with the classical CASENet, we introduce as a new architecture the shearlet Diverse Deep Supervision (Shear-DDS). This architecture will accept as input the shearlet volume of an image. The output consists of the same activation map characterizing the classified edges as the original DDS. 

\begin{figure}[h!]
\centering
\includegraphics[width = 1\textwidth]{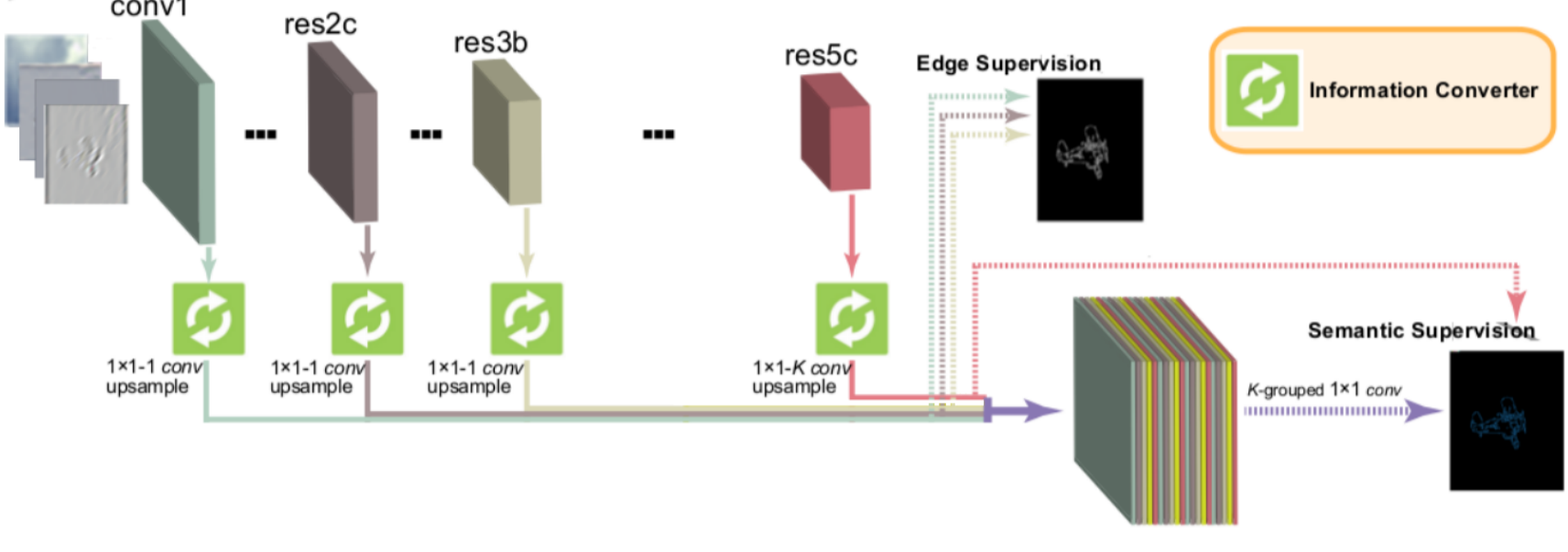}
\caption{Illustration of the shearlet Diverse Deep Supervision architecture (inspired by the figures of \cite{yun2019sed}).}
\label{fig:shearDDS}
\end{figure}

As we did with the Shear-CASENet, we reduced the overall number of parameters by removing stage four of the DDS architecture, with a similar $13\%$ reduction of learnable parameters. Figure~\ref{fig:shearDDS} shows the new Shear-DDS architecture. 

Shear-DDS was trained with the same multi-task loss function as the original architectrue, obtaining an improvement in performance on the SBD dataset, the corresponding numerical results are presented in Section~\ref{subsec:NumResAppsemantic edge detection}.

\section{Numerical Results and Applications}
\label{sec:NumResApp}

To show the true impact of the approach presented in this work, we performed several numerical experiments, targeted to three specific applications, namely, wavefront set extraction, general semantic edge detection, and Computed Tomography reconstruction. 
 In each of the experiments the networks were trained and evaluated on datasets for the specific purpose. On one hand, the general semantic edge detection application was trained and evaluated on the semantic boundaries dataset (SBD), which is already and standard benchmark for this application. 
On the other hand, since wavefront set extraction is well suited to be used in Computed Tomography reconstruction (see Section~\ref{subsec:MicroLocCT}), the DeNSE sub-networks were trained on images formed by random ellipses which resemble human-head phantoms such as the well-known Shepp-Logan phantom.

All experiments show that our hybrid approach, namely combining shearlets with carefully designed network architectures, provide a significant improvement in performance. This indicates that this conceptual approach should be also beneficial for other image processing tasks. 

\subsection{Wavefront Set Extraction}
\label{subsec:NumResWF}

For wavefront set extraction, we trained the DeNSE architecture with the procedure presented in \cite{andrade2019wfset}. We trained the networks on head-like phantom images, inspired by the Shepp-Logan phantom. Our training data consists of a specific selection of random ellipses, two big ellipses, representing the inner and outer skull and small ellipses inside the skull, with different sizes, pixel intensity value, and orientations. We also vary the gradient of the intensity of the ellipses in order to obtain curves with different levels of regularity.

The advantage of using this type of phantom is two-folded: It allows to have access to the analytical wavefront set and the resulting network can be used for Computed Tomography applications in head-like phantoms. 

We used the tensorflow implementation of the DeNSE architecture, publicly available in \url{http://www.shearlab.org/applications}. For the shearlet transform, we used the python interface of the julia API of shearlab, available in \url{http://www.shearlab.org/software}.

The DeNSE architecture classifies patches of an image for an specific given orientation. We used a resolution of 180 distinct orientations and trained the network for each orieantion separately. For each training we used 10,000 images. The evaluation and test was then done over 2,000 images each. We made use of a total of four scales on the shearlet transform, producing a shearlet coefficients volume with 49 slices. From the shearlet coefficients of each image, we extracted 10 distinct patches randomly. This was done in such a way that the classes are balanced, meaning, in each binary classifier, the number of positive cases is roughly the same as the number of negative cases.

We then used the standard MF-score to measure the performance of the classifiers, which is the mean of the F-score over all the orientations. We compared the performance with other semantic edge detection models, and for this made use of the publicly available python code given for the CoShREM \cite{rafael2015coshrem}, CASENet \cite{yu2017casenet} and SEAL \cite{yu2018simultaneous}. For the Yi-Labate-Easley-Krim \cite{yilabate2009shearlet} and the DDS \cite{yun2019sed} models, we used our own implementation. The performance benchmarks with these models are presented in Table~\ref{table:Phantom-MF}.

\begin{figure}[htb!]
\centering
\includegraphics[width = 0.38\textwidth]{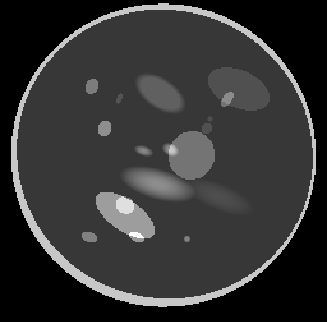}
\includegraphics[width = 0.38\textwidth]{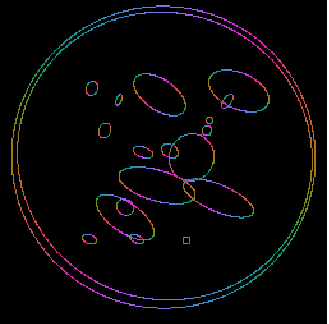} 
\\
\includegraphics[width = 0.38\textwidth]{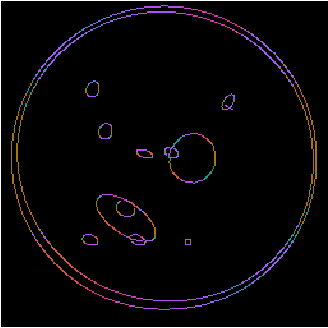}
\includegraphics[width = 0.38\textwidth]{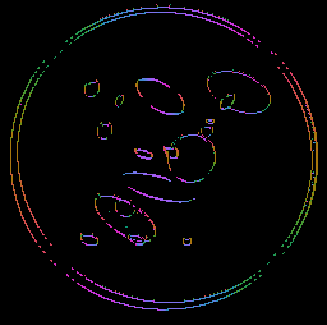}
\\
\includegraphics[width = 0.38\textwidth]{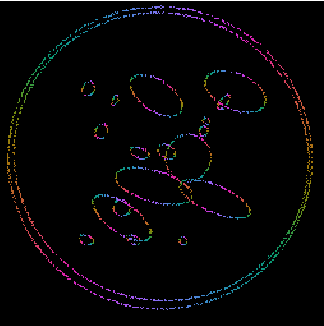}
\includegraphics[width = 0.38\textwidth]{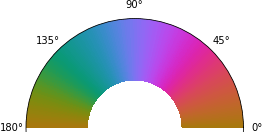}
\caption{Computed edges and orientations of an example of the head-phantom data set. Top-left: Input image. Top-right: Orientations, analytical ground truth. Middle-left: Orientations predicted by CoShREM algorithm. Middle-right:  Orientations predicted by CaSENet. Bottom-left: Orientations predicted by DeNSE\@ algorithm. Bottom-right: Color code for normal-directions.}
\label{fig:WFresults}
\end{figure}

Figure~\ref{fig:WFresults} then shows the results of wavefront set extraction on an example of the head-phantom dataset using three different models, the CoShREM model, and the CASENet architecture. Judging from the images, it is clear that DeNSE shows significantly better performance than CoSHREM, where the latter is not able to find the ellipses with smooth boundaries.  

\begin{table}[htb!]
\centering
\begin{tabular}{l r r r r}
\textbf{Method} & \textbf{MF-score} \\
\hline

Yi-Labate-Easley-Krim\cite{yilabate2009shearlet}  & 75.7\\
\hline

CoShREM\cite{rafael2015coshrem}  & 70.4\\
\hline

CASENet\cite{yu2017casenet} & 78.6 \\
\hline

SEAL\cite{yu2018simultaneous} & 83.4 \\
\hline

DDS\cite{yun2019sed} & 85.6 \\
\hline

\emph{DeNSE}\cite{andrade2019wfset} & \emph{95.7}\\
\hline
\end{tabular}
\caption{Performance of wavefront set extraction on the head-phantom data set. All values are in percentage.}
\label{table:Phantom-MF}
\end{table}

\subsection{General Semantic Edge Detection}
\label{subsec:NumResAppsemantic edge detection}

For the numerical experiments of the general case of semantic edge detection, we trained the Shear-CASENet and Shear-DDS architectures on the Semantic Boundaries Dataset (SBD). This database consists of 11,355 images, from which we used 9,035 images for training, 1,050 for evaluation, and 1,050 for testing. Each image has a human-annotated array of edge-pixels with the intensity value as the category number of the object, where this edge belongs to. The SBD dataset consists of a total of 20 categories, including vehicles, animals, and plants. 

Both Shear-CASENet and Shear-DDS were trained on the full shearlet coefficients. Similar to the case of the wavefront set extraction, we use the digital shearlet transform \cite{kutyniok2012digital} implemented on julia, with a total of four scales. This produces a shearlet coefficients volume of 49 slices, which was then fed to the proposed architectures. 

We use the publicly available implementation of the CASENet architecture (\url{https://github.com/lijiaman/CASENet}). This implementation makes use of the deep learning framework pytorch, making it compatible with our shearlet implementation. Based on this code, we implemented the deep diverse supervision architecture by introducing the information converters and the proposed multi-task loss. Also based on this code, we implemented the Shear-CASENet and Shear-DDS architectures by extending the first convolutional layer with the corresponding shearlet channels (see Figures~\ref{fig:shearDDS} and~\ref{fig:ShearCaseNet}) and removing the fourth stage of the original architectures. 

In addition to CASENet and DDS, we also compared our methods to deeply supervise version of CASENet \cite{yu2017casenet} and SEAL \cite{yu2018simultaneous}. The performance benchmarks presented in Table~\ref{table:SBD-MF} are done in terms of the mean F-score, in a similar fashion as with the wavefront set extraction benchmarks, by computing the mean of the F-score over all the categories. It is visible that the mean-F value is slightly better on the the Shear-CASENet and Shear-DDS than on the other architectures. The improvement is not as significant as in the case of the wavefront set extraction, since DeNSE was specifically designed for this task and the existing models have general semantic edge detection applications. It is though worth to stress that Shear-CASENet and Shear-DDS have significantly less parameters than their non-shearlet counterparts.

\begin{figure}[htb!]
\centering
\includegraphics[width = 0.38\textwidth]{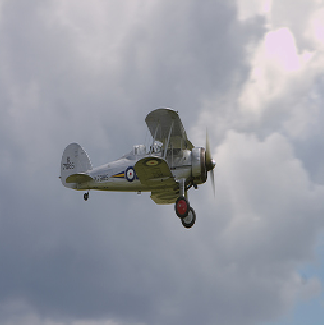}
\includegraphics[width = 0.38\textwidth]{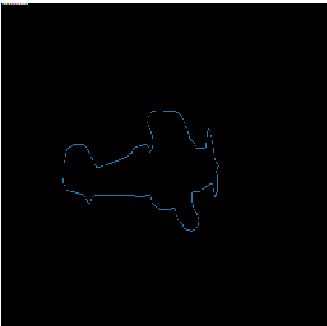} 
\\
\includegraphics[width = 0.38\textwidth]{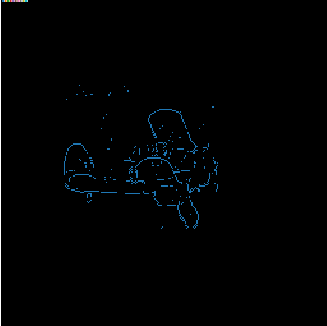} 
\includegraphics[width = 0.38\textwidth]{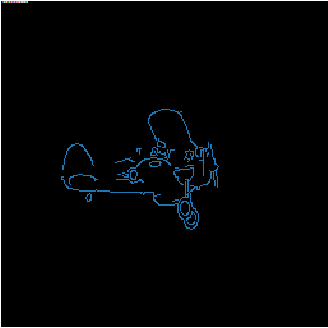}
\\
\includegraphics[width = 0.38\textwidth]{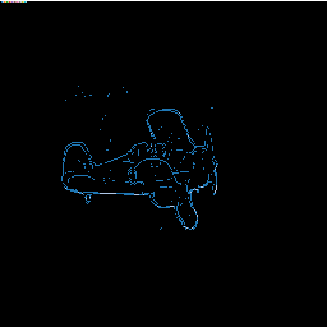}
\includegraphics[width = 0.38\textwidth]{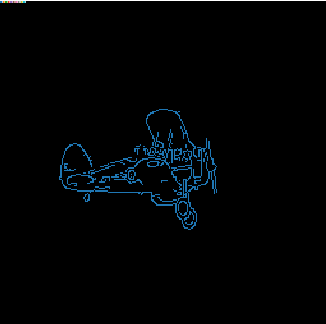}
\caption{Computed semantic edges of an example of the semantic boundaries dataset (SBD). The color blue represents the category of airplane. Top-left: Input image. Top-right: Semantic edges, human annotation. Middle-left: Semantic edges predicted by the classical CASENet architectures. Middle-right: Semantic edges predicted by the classical DDS architecture. Bottom-left: Semantic edges predicted by the Shear-CASENet architecture. Bottom-right: Semantic edges predicted by the Shear-DDS architecture.}
\label{fig:semantic edge detectionresults}
\end{figure}

In Figure~\ref{fig:semantic edge detectionresults}, we depict the results obtained using an example of the SBD dataset. It shows the semantic edges obtained by both CASENet and DDS and their respective shearlet extension. In all the cases the airplane in the picture was correctly classified, but the refinement of the obtained edges is improved in the shearlet version. This strongly suggests that the use of shearlets is well-suited for high performance in semantic edge detection. 

\begin{table}[htb!]
\centering
\begin{tabular}{l r r r r}
\textbf{Method} & \textbf{MF-score} \\
\hline

DSN\cite{yu2017casenet} & 65.2 \\
\hline

SEAL\cite{yu2018simultaneous} & 75.3 \\
\hline

Classical CASENet\cite{yu2017casenet}  & 71.4\\
\hline

Classical DDS\cite{yun2019sed}  & 78.6\\
\hline

\emph{Shear-CASENet}  & 75.7\\
\hline

\emph{Shear-DDS}  & 80.1\\
\hline
\end{tabular}
\caption{Semantic edge detection performance on the SBD dataset. All values are in percentage.}
\label{table:SBD-MF}
\end{table}

\subsection{Tomographic Reconstruction}
\label{subsec:NumResAppCT}

Many inverse problems that arise in tomographic imaging involve recovering object boundaries. It is obvious that determining those boundaries that are visible or not is important. In fact, a close look at the reconstructions reveals that in each case the only feature boundaries that are clearly defined are those tangent to lines in the data set for the problem. Example 1 illustrates this in a simple way: one detects singularities in the Radon data exactly when the lines of integration are tangent to the boundary of the object. The goal of this chapter is to make the idea mathematically rigorous.

As shown in Section~\ref{subsec:MicroLocCT}, microlocal analysis can be used to describe how the Radon transform transforms wavefront sets (singularities). To show this in the context of digitized images and data we make use of the digital wavefront set extraction applied to the head-phantom dataset introduced on Section~\ref{subsec:NumResWF} and the microlocal canonical relation of the radon transform (Section~\ref{subsec:MicroLocFIO}), in order to obtain the digital wavefront set an image from the digital wavefront set of its sinogram without a previous inversion. This approach is based on the work presented by the authors in \cite{andrade2019wfset}. We simulate tomographic data (sinograms) using the python implementation of the digital Radon transform in the operator discretization library (ODL, \url{http://github.com/odlgroup/odl}.

Let us now explain the training procedure.
To label each sinogram with the correct wavefront set, we used a digitized version of the canonical relation for the Radon transform, presented in \cite[Def. 6.1]{andrade2019wfset}. This definition was then taken as an Ansatz for the definition of the digital wavefront set of the sinograms, which came from a phantom whose the wavefront set we know explicitely. We then trained the DeNSE model on the sinogram wavefront sets, with the same training, test, and evaluation set as in Section~\ref{subsec:NumResWF}. Using the results of the canonical relation as the ground truth for the sinogram wavefront sets, we obtained a MF-score over the test datset of $95.7\%$, which is comparable to the performance of DeNSE on the head-phantom image class. We will not present any performance comparison, since we are not aware of any competing algorithm for the detection of wavefront set on sinograms. 

Since applying the inverse canonical relation to the wavefront set of the sinogram will give us access to part of the wavefront set of the original image, without performing yet any inversion, having a method to detect the wavefront set of a sinogram becomes useful for inverse problem regularization.

\begin{figure}[htb!]
\centering
\includegraphics[width = 0.49\textwidth]{figures/figures_Royal/figures_Tomo/phantom_im.png}
\includegraphics[width = 0.49\textwidth]{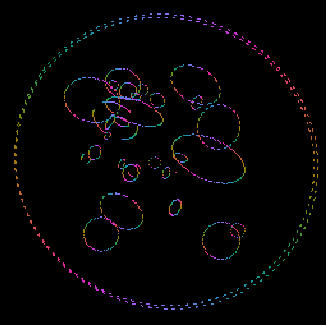}
\\[1em]
\includegraphics[width = 0.49\textwidth]{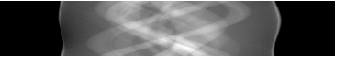}
\includegraphics[width = 0.49\textwidth]{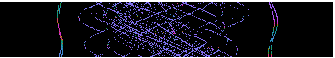}
\\[1em]
\includegraphics[width = 0.38\textwidth]{figures/figures_Royal/cmap.png}
\caption{Top-left: Phantom made from ellipses. Top-right: Associated wavefront set obtained by the inverse canonical relation on the wavefront set of the low-dose sinogram . Middle-left: Low-sinogram, with every six angles measured. Middle-right: Associated wavefront set obtained by DeNSE. Bottom: Color-code for normal directions.}
\label{fig:CTresults2}
\end{figure}

To show this potential, we present in Figure~\ref{fig:CTresults2}, the phantom wavefront set obtained when applying the inverse canonical relation to the low-dose sinogram, when measuring every six degrees. This low-dose problem is highly ill-posed, and will require significant effort to be inverted. Nut having at hand the digital canonical relation will give us almost for free the part of the phantom's wavefront set associated with the measured angles. 

To further show the use of this method, we performed a the low-dose sinogram example, three standard inversion schemes, filtered backprojection, Tikhonov regularization and total variation (TV) regularization. We first compute the associated image reconstruction from the low-dose sinogram, measured each six angles, and then compute the associated wavefront set of the reconstruction with DeNSE. We next compute the average mean square error to the true wavefront set of the data point. Via the inverse canonical relation, we compare this with the error resulting from computing the wavefront set of the sinogram and mapping it back to the image. 

\begin{figure}[htb]
\centering
\includegraphics[width = 0.38\textwidth]{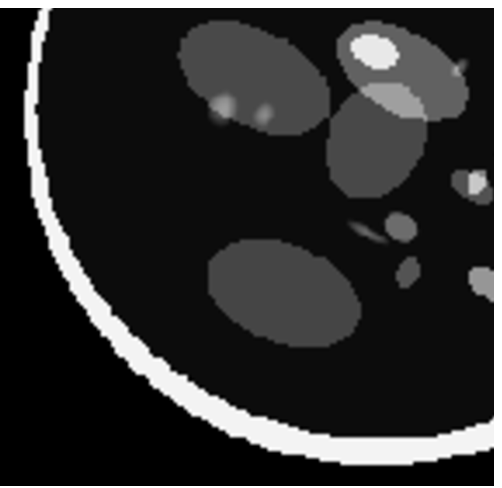}
\includegraphics[width = 0.38\textwidth]{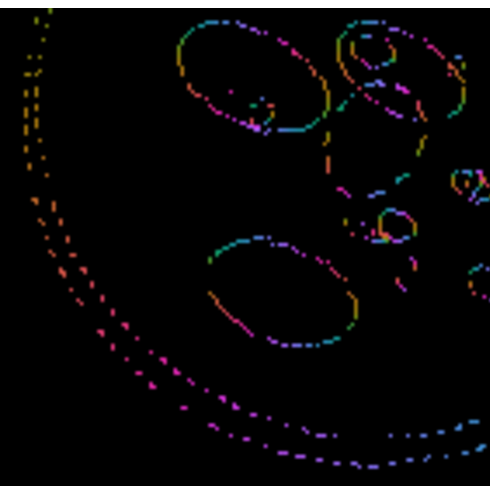}
\\
\includegraphics[width = 0.38\textwidth]{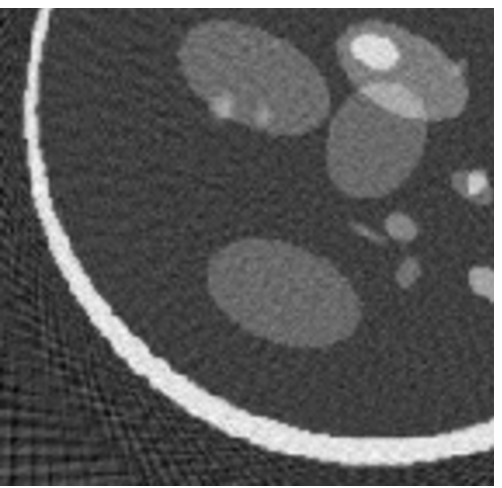}
\includegraphics[width = 0.38\textwidth]{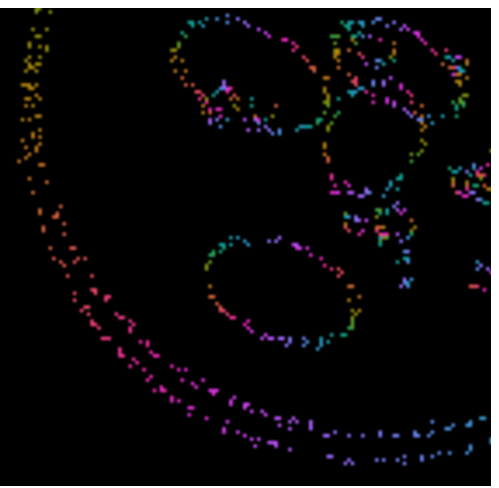}
\\ 
\ \includegraphics[width = 0.38\textwidth]{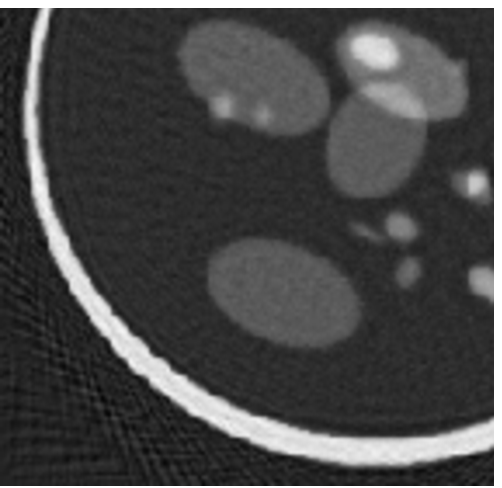}
\includegraphics[width = 0.38\textwidth]{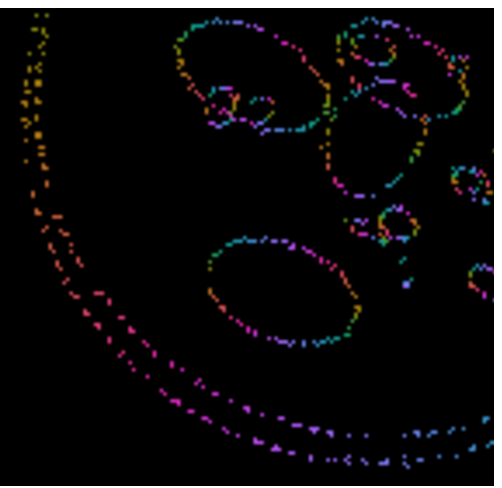}
\caption{Top-left: Phantom made from ellipses. Top-right: wavefront set of the phantom data computed by the inverse canonical relation on the low dose sinogram wavefront set extracted by DeNSE. 
Middle-left: Filtered back projection reconstruction. Middle-right: wavefront set of the filtered backprojection reconstruction extracted by DeNSE. 
Bottom-left: Tikhonov reconstruction. Bottom-right: wavefront set of the Tikhonov reconstruction extracted by DeNSE.}\label{fig:CTResults3}
\end{figure}

Figure~\ref{fig:CTResults3} shows the wavefront sets corresponding to the different reconstruction schemes. In Table~\ref{table:TV-results}, the obtained errors are presented, clearly showing the advantage of first extracting the wavefront set of the sinogram and then applying the canonical relations, over any first-invert-then-extract strategy. In additio,n this a-priori information can be used as a regularizer on any variational regularization scheme.

\begin{table}[htb!]
\centering
\begin{tabular}{l r r r r}
\textbf{Inversion technique} & \textbf{Mean square error}\\
\hline
Tikhonov & 443.0\\
\hline

Total variation & 380.9\\
\hline

Filtered backprojection & 504.3 \\
\hline
Canonical relations & 168.1 \\
\hline
\end{tabular}
\caption{Error of wavefront set estimation by different inversion techniques.}
\label{table:TV-results}
\end{table}

\begin{figure}[htb!]
\centering
\includegraphics[width = 0.49\textwidth]{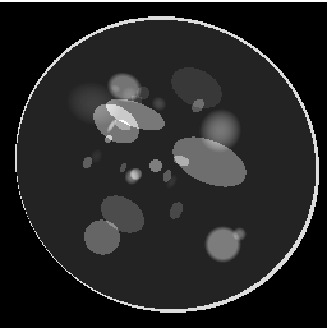}
\includegraphics[width = 0.49\textwidth]{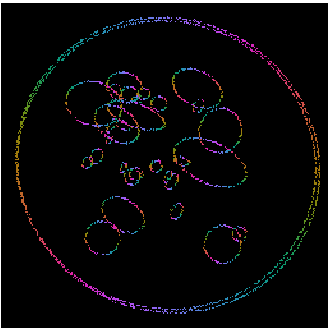}
\\[1em]
\includegraphics[width = 0.49\textwidth]{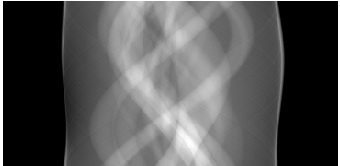}
\includegraphics[width = 0.49\textwidth]{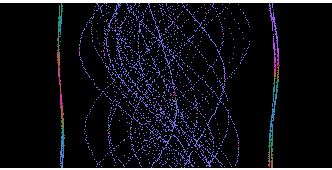}
\\[1em]
\includegraphics[width = 0.38\textwidth]{figures/figures_Royal/cmap.png}
\caption{Top-left: Phantom made from ellipses. Top-right: Associated wavefront set extracted by DeNSE. Middle-left: Radon transform of the phantom. Middle-right: Associated wavefront set computed through digital canonical relations. Bottom: Color-code for normal directions.}
\label{fig:CTresults1}
\end{figure}

\subsection{Discussion of Numerical Experiments}
\label{subsec:NumResConcl}

We observed in the numerical experiments an improvement on edge extraction and classification performance when using the shearlet transform for model-based feature extraction previous to feeding the network. 
In particular, the DeNSE model for wavefront set extraction has shown a significant improvement with respect of traditional methods designed for wavefront set extraction (\cite{rafael2015coshrem}, \cite{yilabate2009shearlet}), due to the combination of both the model-based shearlet representation and the data-driven high performance classification. On the other hand, it also presents better performance with respect of deep-learning based models (\cite{yu2017casenet}, \cite{yu2018simultaneous}, \cite{yun2019sed}), although is worth to mention that this methods were originally designed for semantic edge detection with a lower variation in the classes over the contiguous pixels. In addition DeNSE, by definition, is able to avoid successfully the distracted supervision paradox, without much effort.

For general semantic edge detection, introducing a change of representation system, to the shearlet system, improved the performance of the standard semantic edge detection models (CASENet and DDS), but not at the same level. This suggests, that these models are not well suited to the problem of wavefront set extraction, and a patch-based strategy will reach to a similar performance leap as in the wavefront set extraction case. 

The results also suggest that the use of shearlets in image processing tasks involving edge detection, helps to do heavy-lifting in a model-based fashion, reducing the amount needed parameters, resulting in a reduction of complexity. This is clearly shown by the shear-CASENet and shear-DDS architecture which perform better than their non-shearlet version, but have less learnable parameters. This combination of model-based and data-driven approaches is a strategy that more researchers have adopted in the last few years. 

In addition, the applications of semantic edge detection in inverse problems, are not yet explored in depth, but the results presented on Section~\ref{subsec:NumResAppCT}, are a clear indicative that this area will be fruitful in the next few years. As part of the future developments, the authors will work on the use for the wavefront set extractor to improve the existing hybrid methods on tomographic reconstruction.

\enlargethispage{20pt}




\section*{Acknowledgements}
H.A.-L. is supported by the Berlin Mathematical School and MATH+. G.K. acknowledges partial support by the Bundesministerium f\"ur Bildung und Forschung (BMBF) through the Berliner Zentrum for Machine Learning (BZML), Project AP4, RTG DAEDALUS (RTG 2433), Projects P1 and P3, RTG BIOQIC (RTG 2260), Projects P4 and P9, and by the Berlin Mathematics Research Center MATH+, Projects EF1-1 and EF1-4. The work of O.\"O. was supported by the Swedish Foundation of Strategic Research grant AM13-004. 




\vskip2pc

\bibliographystyle{abbrv}
\bibliography{references}

\end{document}